\documentclass[preprint]{aastex}

\shorttitle{OBLIQUITY TIDES}

\shortauthors{FABRYCKY, JOHNSON, \& GOODMAN}

\usepackage{amsmath,amssymb,natbib,color,lscape}

\renewcommand{\b}[1]{\boldsymbol{#1}}

\newcommand{\bm}[1]{\mbox{\boldmath{$#1$}}}

\newcommand{\unit}[1]{\,{\rm #1}}

\newcommand{\um}{\,\mu{\rm m}}

\newcommand{\erg}{\unit{erg}}

\newcommand{\K}{\unit{K}}

\newcommand{\msun}{\unit{M_\odot}}

\newcommand{\s}{\unit{s}}

\newcommand{\yr}{\unit{yr}}

\newcommand{\bdot}{\b{\cdot}}

\newcommand{\cross}{\b{\times}}

\newcommand{\Shat}{\mathbf{\hat S}}

\newcommand{\rhat}{\mathbf{\hat r}}

\newcommand{\xhat}{\mathbf{\hat x}}

\newcommand{\yhat}{\mathbf{\hat y}}

\newcommand{\zhat}{\mathbf{\hat z}}

\newcommand{\Xhat}{\mathbf{\hat X}}

\newcommand{\Yhat}{\mathbf{\hat Y}}

\newcommand{\ehat}{\mathbf{\hat e}}

\newcommand{\qhat}{\mathbf{\hat q}}

\newcommand{\hhat}{\mathbf{\hat h}}

\newcommand{\Ophat}{\mathbf{\hat{\Omega}_p}}

\newcommand{\Jhat}{\mathbf{\hat J}}

\newcommand{\J}{\mathbf{J}}

\newcommand{\tidalQ}{Q}
\newcommand{\tidalQp}{Q'}

\begin{document}


\slugcomment{Submitted to ApJ}

\title{Cassini States with Dissipation: Why Obliquity Tides Cannot
  Inflate Hot Jupiters}

\author{Daniel C. Fabrycky, Eric T. Johnson, and Jeremy Goodman}

\affil{Department of Astrophysical Sciences,
 Princeton University, Princeton, NJ 08544}
 
\begin{abstract}
  Some short-period exoplanets (``hot Jupiters'') are observed by
  their transits to have anomalously large radii.  It has been
  suggested that these planets are in a resonance involving persistent
  misalignment and synchronous precession of their spin and orbital
  angular momenta---a Cassini state---and that the attendant tidal
  heating inflates the planet.  We argue against this.  Using explicit
  tidal integrations, we show that although an oblique Cassini state
  can dissipate many times the planet's rotational energy, the
  \emph{rate} of dissipation must be much less than hypothesized.
  Dissipation causes the planetary spin to lie at an
  angle to the plane containing the orbital and total angular momenta.
  If dissipation is too rapid, this angle becomes so large that Cassini
  equilibrium is lost.  A separate consideration limits the total
  energy that can be extracted from the orbit.  The source of the
  torque on the orbit, either an oblique parent star or an inclined
  third body, aligns with the orbit as it absorbs the angular
  momentum shed by the planet.  Alignment removes the orbital
  precession required by the Cassini state.  In combination with
  observational bounds on the mass and semimajor axis of a possible
  second planet and with bounds on the stellar rotation and obliquity,
  these constraints make it very unlikely that obliquity tides can be
  the explanation for inflated hot Jupiters, especially HD~209458b.
\end{abstract}


\section{Introduction}

Transit observations have exposed the limits of our understanding of
the structure of short period extrasolar giant planets, or hot
Jupiters.  The models of transiting planet HD~209458b constructed by
\citet*{Bodenheimer_etal01} gave the first indication of a discrepancy
between the theoretically predicted and observed radii of the largest
hot Jupiters.  While standard evolutionary models for irradiated giant
planets faithfully represent the majority of the 14 known transit
cases, the conclusion persists that the models underpredict the radii
of several hot Jupiters by $10-20\%$
\citep{Burrows_etal06a,Arras_Bildsten06,Fortney_etal06b,Guillot05}.  At
the large end of the size distribution, planets HD~209458b
\citep{Knutson_etal07}, WASP-1b \citep{Shporer_etal06}, and HAT--P-1b
\citep{Bakos_etal06} present the most serious challenge to expected
rates of contraction.  Resolution of the apparent radius problem will
require simultaneous treatment of several outstanding issues, with
competing effects.  First, the problem is ameliorated somewhat by
distinguishing the photospheric radius from the ``transit radius,''
which is the impact parameter of photons that travel a chord of unit
optical depth through the planetary atmosphere.  The latter radius is
larger by as much as $\sim 5\%$ for planets with low surface gravity
\citep{Baraffe_etal03,Burrows_etal03,Burrows_etal06a}.  Secondly,
significant enlargement can be achieved in two ways; either by tapping
an additional heat source \citep{Bodenheimer_etal01} or by insulating
the interior with a supersolar abundance of metals in the atmosphere,
thereby retarding its contraction \citep{Burrows_etal06a}.
Conversely, more rapid contraction may result if global redistribution
of the insolation is inefficient, allowing an increased cooling flux
to escape via the planet's cold night side \citep{Guillot_Showman02}.
Finally, planet radii are greatly reduced by the inclusion of heavy
elements in the interior, whether dispersed or as a solid core, for
which a mass $10-40$ times the mass of the Earth may be typical
\citep{Bodenheimer_etal03,Fortney_etal06b,Burrows_etal06a}.

\citet{Burrows_etal06a} argue that, to within the observational
uncertainties in radius and age, with no additional heating mechanism,
the radii of all transiting hot Jupiters are consistent with
evolutionary models if supersolar atmospheric metallicities are
allowed.  The authors rely upon heterogeneity in both atmospheric
abundances and core mass.  In particular, Burrows et al.~fit coreless
models with 10 times solar atmospheric abundances to WASP-1b (within
the 1 $\sigma$ error bar), HAT--P-1b (within 1 $\sigma$), and
HD~209458b (within 2 $\sigma$).  A concern is that the inflated radii
of these models require that the metals be sequestered in the
atmosphere; supersolar abundances throughout the planet might be
expected to result in a net contraction \citep{Fortney_etal06c}.
Observations of thermal emission from hot Jupiters also bear on the
relevance of such models.  Broadband detections of infrared emission
at secondary eclipse have been reported for HD~209458b
\citep{Deming_etal05}, TrES-1 \citep{Charbonneau_etal05}, and
HD~189733b \citep{Deming_etal06}.  Corresponding theoretical spectra
were constructed by \citet{Seager_etal05}, \citet{Barman_etal05},
\citet{Fortney_etal05}, and \citet{Burrows_etal05,Burrows_etal06b}.
All obtained fits to the emission measurements while assuming
\emph{solar} abundances.  None predicted metallicity as large as
required by \citet{Burrows_etal06a} for HD~209458b, although
\citet{Fortney_etal05} found marginally improved fits for TrES-1 using
$3-5$ times solar abundances.  Interpretation should nonetheless be
cautious, as the continua of the model spectra depend more strongly on
the degree of horizontal heat transport than on metallicity.  In fact
there is no consensus among these works as to whether the data
indicate that the reradiation of absorbed stellar flux occurs locally,
is globally uniform, or is intermediate between these limits.  This
points to another critical issue; insofar as the photospheric
temperature is an indicator of the temperature near the base of the
radiative zone, the observed day-night temperature contrast can
constrain the redistribution of heat at depth.  The contraction rates
computed by Burrows et al. hinge on thorough redistribution and
isothermal conditions along the radiative-convective boundary, where
the cooling flux is determined \citep{Arras_Bildsten06}.  Such an
isotherm bounding the convective interior was seen in the numerical
simulations of global circulation by \citet{Iro_etal05}.  However, if
this deep isotherm does not exist, then additional heating is indeed
required to offset the steeper temperature gradient on the night side.
\citet{Harrington_etal06} reported measurements of the phase curve of
(non-transiting) hot Jupiter $\upsilon$~Andromedae~b that imply a
day-night contrast perhaps exceeding $10^3\K$.  However, from
simulations with a deep isotherm, Iro et al. predicted contrasts not
larger than $600\K$ for HD~209458b (despite the fact that HD~209458b
receives greater stellar flux than $\upsilon$~And~b).  As a final
point, the first spectroscopic measurements of thermal emission have
come with their own surprises.  The mid-infrared spectra of HD~189733b
\citep{Grillmair_etal07} and HD~209458b
\citep{Richardson_etal07,Swain_etal07} are consistent with a
Rayleigh-Jeans continuum, lacking the anticipated absorption bands
blueward of $9\um$ associated with water and methane.  These
observations of hot Jupiter atmospheres will need to be understood
theoretically before a conclusion can be reached regarding heat
retention.  A continual heating mechanism may yet play a principal
role.

Augmentation of a Jovian planet's intrinsic luminosity causes an
adjustment to a new hydrostatic equilibrium with an expanded radius.
The estimated heating rates necessary to account for the inflated hot
Jupiters are in the range $10^{26}-10^{28}\erg\s^{-1}$, which is
$\lesssim 5\%$ of the intercepted stellar luminosity and is a few
orders of magnitude larger than the intrinsic cooling luminosity
\citep{Burrows_etal06a,Bodenheimer_etal03,Bodenheimer_etal01,Baraffe_etal03}.
This energy may be drawn from either the stellar insolation or the
orbit.  \citet{Guillot_Showman02} pointed out that one-sided
irradiation should drive an atmospheric heat engine that converts the
insolation into bulk flow kinetic energy.  Inflation would require
that a fraction of the kinetic energy be dissipated at depths where
the radiative timescale is sufficiently long.  Numerical global
circulation models that address this mechanism have yet to converge on
several key factors, including characteristic wind velocities and
global flow patterns.  While \citet{Showman_Guillot02} found
substantial downward transport of kinetic energy in simulation,
\citet{Burkert_etal05} found little vertical transport and near
complete dissipation at shallow depths.  A further puzzle is that
\citet{Harrington_etal06} observed the phase curve of $\upsilon$~And~b
to be consistent with an $11^\circ$ \emph{lag} between full phase and
the substellar point.  A superrotational wind like that predicted by
\citet{Showman_Guillot02} and \citet{Cooper_Showman05} for HD~209458b,
with a $60^\circ$ \emph{advance}, is strongly disfavored by the data.
Presuming that dissipation does occur at sufficient depth to affect
the radius, one has to explain why this mechanism---or any other that
appeals to characterstics common to all hot Jupiters---does not
inflate \emph{all} currently observed transiting planets.  A possible answer involves competition
among conflicting factors: for example, auxiliary heating offset by
variable core masses.

An alternative is that tidal forces may heat the planet at the expense
of the orbit.  Tidal dissipation naturally depends upon variable
orbital and perhaps other parameters.  The action of tides is to align
the spin and orbital angular momenta, synchronize the rotational and
orbital periods, and circularize the orbit.  The efficiency with which
hot Jupiters are thought to dissipate tidal oscillations implies that
alignment and synchronization occur on $10^5-10^6\yr$ timescales and
circularization occurs on $10^8-10^9\yr$ timescales.  The eccentricity
and/or obliquity must be excited continually in order for tides to
persist for several Gyr.  Unlike the case of insolation-driven
heating, ongoing tidal heating requires special orbital
configurations, potentially involving a third body.
\citet{Bodenheimer_etal01} and \citet{Bodenheimer_etal03} have
suggested that eccentricity could be forced by a companion planet, and
they have estimated that a sustained eccentricity $e\gtrsim 0.03$
would be sufficient to explain the radius of HD~209458b.  Recent
radial velocity measurements \citep{Laughlin_etal05b} and timings of
secondary eclipse \citep{Deming_etal05} have tightened the constraints
on the eccentricity of this system, but because of degeneracy with the
longitude of periastron, cannot yet rule out $e=0.03$ with $95\%$
confidence.  An innovative possibility, put forth by
\citet{Winn_Holman05}, is that a planet could occupy a so-called
Cassini state in which its obliquity, rather than its eccentricity, is
forced.

G. D. Cassini published three empirical laws in 1693 describing the
spin state of the moon.  Cassini's laws state that (1) the moon's
rotational and orbital periods are synchronous, (2) its equatorial
plane maintains constant inclination to the plane of the ecliptic, and
(3) its spin axis remains coplanar with the normal to its orbit and
the normal to the ecliptic.  The first law is an outcome of tidal
evolution for bodies with a permanent quadrupolar distortion, like the
moon, or for fluid bodies in circular, stationary orbits
\citep{1999MD}.  The second and third laws characterize a Cassini
state, which is an equilibrium for the motion of the spin axis of a
body in a uniformly precessing Keplerian orbit.  Up to four Cassini
states may exist for a given system, each with a preferred obliquity.
The moon occupies Cassini state 2 (see numbering scheme in
\S\ref{sec:review}) with obliquity $\theta\approx 6\fdg7$.  Resonance
between the precession frequency of the orbit and the natural
precession frequency of the spin determines the values of the Cassini
obliquities.  The obliquity of a dissipative body is driven toward a
Cassini state, which is a quasi-equilibrium in the sense that the body
continues to experience an obliquity tide so long as $\theta$ is
nonzero.  In this paper we determine the effect of the attendant
obliquity tide on the evolution and survival of Cassini states for hot
Jupiters.

\citet{Levrard_etal07} have recently argued that the probability of a
planet becoming trapped in an oblique Cassini state is rather small.
Such probabilities depend upon the early evolution of the star and
perhaps of its protostellar disk. Independently of initial conditions,
however, there are reasons to doubt whether a hot Jupiter can
\emph{remain} in Cassini state 2 while experiencing ongoing tidal
heating sufficient to affect the planet's radius measurably.
\citet{Levrard_etal07} themselves have pointed out that maintenance of
the oblique Cassini states requires a balance between the torque
associated with dissipation in the planet and a nondissipative torque
on the orbit due either to an oblate and oblique stellar primary or to
a third body.  Since the nondissipative torque on the orbit is likely
to be small, we argue that the Cassini state can be maintained only by
assuming a dissipation rate in the planet that is too small to inflate
its radius significantly.  Also, \citet{Levrard_etal07} considered
only the balance of torques tending to alter the obliquity.  Using a
specific tidal model, we show that a stronger constraint results from
requiring the spin to equilibrate in the azimuthal direction.

\begin{figure}
\plotone{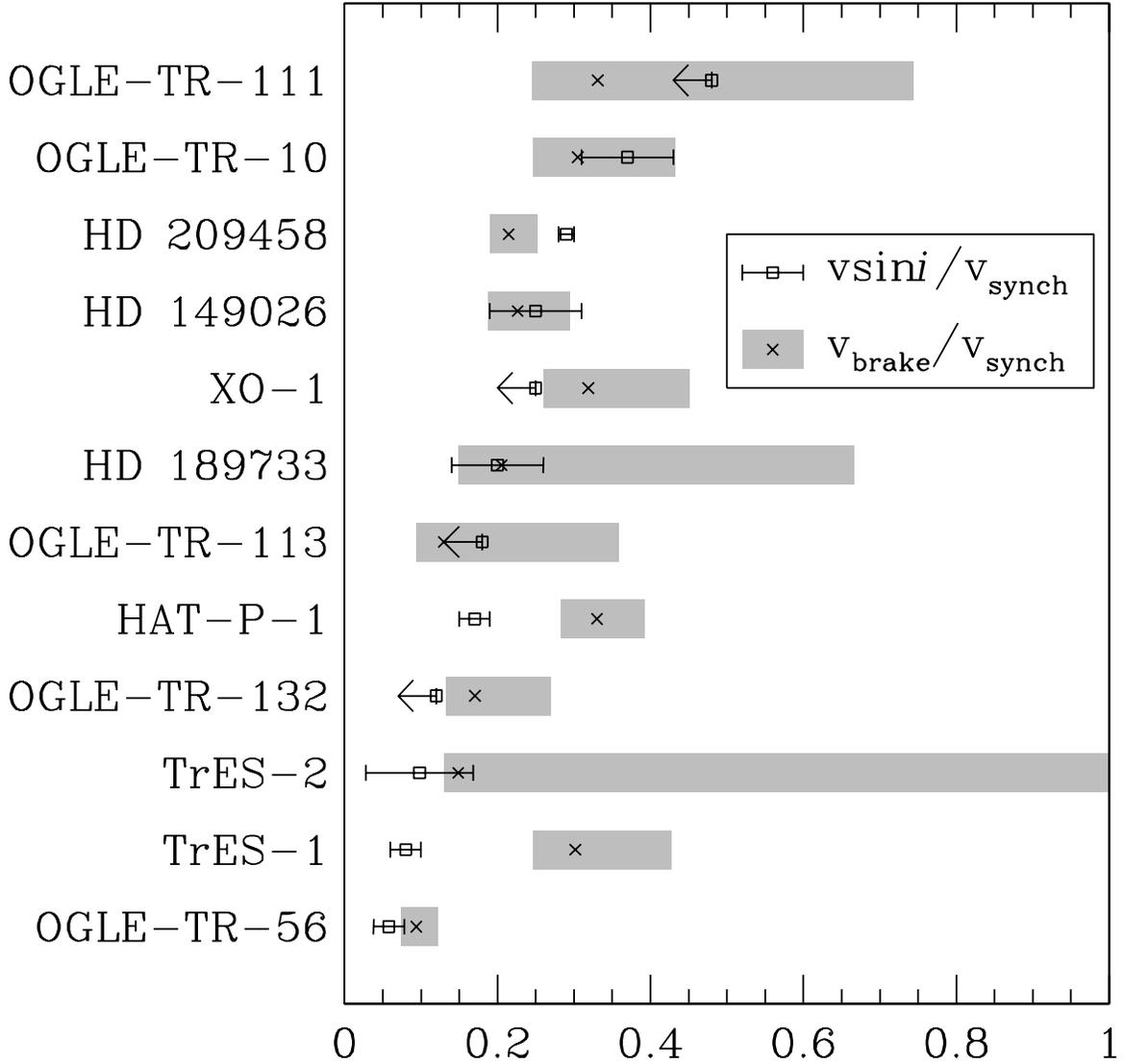}
\caption{Measured stellar rotational velocity $(v\sin i)_\star$
relative to synchronous rotation $v_{\rm synch}=nR_\star$ (squares).
Stars bearing transiting planets are listed top to bottom in order of
decreasing $v\sin i/v_{\rm synch}$.  Error bars indicate observational
uncertainty in $(v\sin i)_\star$ and $R_\star$.  We compute $v_{\rm
brake}$ from a Skumanich law (see text) and plot $v_{\rm brake}/v_{\rm
synch}$ (crosses).  Scatter in $v_{\rm brake}$ is $\approx 10\%$.
Gray boxes give a range for $v_{\rm brake}/v_{\rm synch}$
corresponding to the 1 $\sigma$ uncertainty in stellar age (systematic
uncertainties are not accounted for).  Probably all stars are rotating
subsynchronously, since $v\sin i\approx v$ for a transiting planet
whose orbit is nearly coplanar with the equator of its host star.
Only HD~209458 is rotating (marginally) more rapidly than would be
predicted by magnetic braking alone.  Data are taken from
\citet{Burrows_etal06a} and references therein, except for the values
of $(v\sin i)_\star$, which are taken from the following sources, top
to bottom: \citet{Pont_etal04}, \citet{Melo_etal06},
\citet{Winn_etal05}, \citet{Sato_etal05}, \citet{McCullough_etal06},
\citet{Bouchy_etal05}, \citet{Bouchy_etal04}, \citet{Bakos_etal06},
\citet{Bouchy_etal04}, \citet{O'Donovan_etal06},
\citet{Laughlin_etal05a}, and \citet{Melo_etal06}.}
\label{fig:skumanich}
\end{figure}

Our investigation, however, was stimulated not by the points above but
by concerns about the orbital angular momentum.  As noted by
\citet{Winn_Holman05}, the suggested tidal input of $\gtrsim 4\times
10^{26}\,{\rm erg\,s^{-1}}$ required to explain HD~209458b's present
radius \citep{Bodenheimer_etal03}, if continued for $5\,{\rm Gyr}$, is
twice the present orbital energy of the planet.  Hence, even if a
somewhat smaller tidal power were acceptable or the planet suffered
less dissipation in the past, the orbital energy must have changed by
of order its present value, with a large loss of orbital angular
momentum.  \citet{Winn_Holman05} have not explained where the angular
momentum might have gone.  It is not likely to have been absorbed by
the star, for two reasons.  First, the moment of inertia of the star
is somewhat smaller than that of the orbit (\S\ref{subsec:tidalQs}),
so that, notwithstanding magnetic braking by a stellar wind, stars
hosting inspiraling hot Jupiters might be expected to rotate more
rapidly than other stars of similar age and spectral type, contrary to
the evidence of Figure~\ref{fig:skumanich} in this paper.  Second, the
timescale for dissipating the planetary tide in the star probably
exceeds its age (\S\ref{subsec:tidalQs}), which limits the amount of
angular momentum that the star can absorb.  Without dissipation, the
star can change only the direction but not the magnitude of its spin
angular momentum.\footnote{In the equilibrium tide, the tidal bulge
raised in the star by the planet points directly at the planet, since
the phase lag between the bulge and the perturber is proportional to
the dissipation rate.  Hence the planet exerts no torque on the tidal
bulge, while any torque exerted on the star's rotational oblateness is
perpendicular to the spin axis and therefore changes only the
direction of the spin.  Alternatively---and the following argument
applies to dynamical as well as equilibrium tides---we may apply
Kelvin's Circulation Theorem to a closed contour drawn along the
rotational equator.  The circulation $2\pi R_\star v_{\rm rot}$ is
conserved in a nondissipative star, since $(\b{\nabla\times
v})\b{\cdot\nabla}S$ is conserved following the fluid and we may take
the curve to be the boundary of a surface of constant specific entropy
$S$.  A contradiction would therefore result if the equatorial
velocity were to evolve secularly without dissipation.}  Thus, the
amount of angular momentum that the star can absorb from the orbit is
limited by the star's initial obliquity as well as its moment of
inertia.  In four cases where the spectroscopic transit
(Rossiter-McLaughlin effect) has been well measured, the implied angle
($\lambda$) between the projections of the stellar spin and the
orbital angular momentum onto the plane of the sky is small \citep[and
references therein]{Winn06}; in particular, $\lambda=-4\fdg 4\pm 1\fdg
4$ for HD~209458 \citep{Winn_etal05}.  Unless the stellar spin has a
large component along the line of sight---which has been ruled out for
HD~189733 at least \citep{Winn_etal06a}---the smallness of $\lambda$
indicates that if maintenance of the Cassini state relies upon angular
momentum changes in the star, then the future life expectancy of that
state is much less than the present age of the system.

The plan of this paper is as follows.  \S\ref{sec:review} presents a
pedagogical review of Cassini states in general nondissipative,
Hamiltonian systems.  \S\ref{sec:tides} discusses causes of the
precession of the orbit in more specificity: either an oblate and
oblique primary, or a third body.  Tidal dissipation is introduced,
and it is argued from empirical constraints on tidal quality factors
that dissipation in the planet is more important than dissipation in
the star for this problem.  \S\ref{sec:numerical} presents and
interprets the governing equations used to describe the tidal
evolution of the planetary spin and orbit, which are adopted from
\cite{2001EK} with small notational changes.  In
\S\ref{sec:cassini_evolve}, we present exemplary numerical results and
interpret them using analytic approximations.  We explore how Cassini
state 2 ``breaks'' due to the imbalance of dissipative and
non-dissipative torques, and we find a strong constraint on the
planetary dissipation rate that renders obliquity tides unimportant as
an internal heat source.  In \S\ref{sec:discussion}, we summarize our
main results and use them to quantify the objections raised above to
Cassini states as an explanation for the anomalously large radius of
HD~209458b in particular.

\section{Cassini states}\label{sec:review}

The spin dynamics of a body in a precessing Keplerian orbit are
analogous to orbital dynamics in a rotating 2D potential.  In both
systems a corotating frame is found in which the stationary points of
the effective potential are equilibria.  The spin counterparts to
Lagrange points are Cassini states.  In this section we review the
theory of Cassini states.

The modern study of spin equilibria was begun by \citet{Colombo66},
who generalized Cassini's laws for axisymmetric bodies.
\citet{Peale69} first enumerated the Cassini states and generalized to
the non-axisymmetric case.  Initial applications of the theory were
made primarily to Mercury and the moon, but also to Iapetus and Triton
\citep{Colombo66,Peale69,Peale74,Beletskii72,Ward75}.  Evolution of
the obliquity of Mars due to geophysical changes in the planet's
oblateness was studied by \citet{Ward_etal79} and
\citet{Henrard_Murigande87}.  The latter work contains a particularly
illuminating development of the Cassini formalism.  Recent
applications to Saturn \citep{Ward_Hamilton04} and Jupiter
\citep{Ward_Canup06} suggest that the primordial obliquities of many
solar system bodies may have been altered by trapping in Cassini
resonances.  Our present concern is with application to hot Jupiters
\citep{Winn_Holman05,Levrard_etal07}.  The essential features of
Cassini states may be derived by specializing to the case of a
dissipationless, axisymmetric body.  For a general treatment see
\citet{Peale74}.  We discuss the critical role of tidal dissipation in
\S\ref{sec:cassiniwdiss}.

Consider an axisymmetric planet with principal moments of inertia
$C>A=B$ in orbit around a star of mass $M_\star$.  Let the total
angular momentum of the system (including other bodies) be denoted
$\J$, pointing along the fixed unit vector $\Jhat$.  This axis and the
invariable plane perpendicular to $\Jhat$ are at rest in an inertial
frame.  Let the orbital plane be inclined to the invariable plane by
angle $I$, and allow the node to precess uniformly around $\Jhat$ with
angular frequency $g$ due to an unspecified constant torque.  In the
two-body problem the nodal precession is dominated by the stellar
quadrupole, but torques from additional bodies or the circumstellar
disk may dominate when they are present.  The orbit normal $\hhat$ remains
fixed in a frame rotating with angular velocity $\mathbf g=g\Jhat$
with respect to the inertial frame.  In cases of interest the node
regresses ($g<0$).  Define a Cartesian coordinate system $XYZ$
attached to the rotating frame as in Figure~\ref{fig:coords}, with the
$Z$-axis along the orbit normal (${\bf\hat{Z}} = \hhat$) and the
$X$-axis in the direction of the ascending node.  The orientation of
the planet relative to the rotating frame is specified by the usual
Euler angles $(\theta,\phi,\psi)$, which give the positions of the
planet's axis of symmetry $z$ and the arbitrary orthogonal axes $x$
and $y$ frozen into the body.  The equatorial plane intersects the
orbital plane at an obliquity $\theta=\cos^{-1}(\zhat\bdot\hhat)$.
This node precesses around $\hhat$ with angular frequency $\dot\phi$.
The planet rotates with respect to the inertial frame with
instantaneous angular velocity $\bm\omega$, which is the sum of the
spin angular velocity $\mathbf\Omega\equiv\dot\psi\zhat$ and the
angular velocity associated with the precession of the symmetry axis.
If the $z$-axis is at rest in the rotating frame then
$\bm\omega=\mathbf\Omega+\mathbf g$.  The planet's total spin angular
momentum $\mathbf S=C\omega_z\zhat+B\omega_y\yhat+A\omega_x\xhat$ can
be approximated by principal axis rotation if the precessional angular
frequency is small compared to $\Omega$, in which case only the
$\zhat$ component of $\mathbf S$ is retained.  The further
approximation $\omega_z\approx\Omega$ is often made for
simplification, and we do so here.  In Figure~\ref{fig:coords} we have
implicitly assumed principal axis rotation by taking $\mathbf
S=C\mathbf\Omega$.  The Hamiltonian $\mathcal H_{\rm RF}$ governing
the spin of the planet in the rotating frame is related to the
Hamiltonian $\mathcal H_{\rm IF}$ in the inertial frame by $\mathcal
H_{\rm RF}=\mathcal H_{\rm IF}-\mathbf S\bdot\mathbf g$.  This is
simply the relation between Jacobi's integral of motion in a rotating
potential and the total energy.  $\mathcal H_{\rm IF}$ is the total
energy associated with the planet's orientation, and $\mathcal H_{\rm
RF}$ is an integral of motion for constant $\mathbf g$ and in the
absence of dissipation.  Hereafter we drop the subscript RF.  The
Hamiltonian in the rotating frame
\begin{equation}\label{H}
\mathcal H=\frac{S^2}{2C}+V-\mathbf S\bdot\mathbf g
\end{equation}
includes the orientation-dependent component $V$ of the planet's
gravitational potential.  This is contributed by spin-orbit coupling,
while spin-spin coupling by the stellar quadrupole is negligible.
Expanded in spherical harmonics and truncated at second order,
\begin{equation}\label{V}
V=\frac{GM_\star(C-A)}{r^3}\left[\frac{3}{2}(\rhat\bdot\Shat)^2-\frac{1}{2}\right],
\end{equation}
where $\rhat$ is the instantaneous direction of the planet with
respect to the star and $r$ its distance.  $V$ has a short-term
variation over the orbit and a long-term variation with the spin
precession.  We may average $V$ over an orbital period while holding
$\Shat$ constant to obtain the mean potential on timescales much
greater than the orbital period $P$.  For an orbit with mean motion
$ n = 2\pi/P$ and eccentricity $e$,
\begin{equation}\label{Vavg}
\frac{1}{P}\int_0^P\!Vdt=-\frac{n^2(C-A)}{2(1-e^2)^{3/2}}\left[\frac{3}{2}(\Shat\bdot\hhat)^2-\frac{1}{2}\right],
\end{equation}
which depends only on $\Shat$ since $\hhat$ is fixed in the rotating
frame.  The system described by (\ref{H}) and (\ref{Vavg}) has 2
degrees of freedom.  Its state may be specified by the generalized
coordinates $(\psi,\phi)$ and conjugate momenta $(S,S\cos\theta)$.  It
is immediately apparent that $S$ is conserved due to the axisymmetry
of the planet.  We define the constants
\begin{equation}\nonumber
\alpha\equiv\frac{3 n^2(C-A)}{2C\Omega(1-e^2)^{3/2}}\qquad\mbox{and}\qquad\mathcal
H'\equiv\left(\mathcal H-\frac{S^2}{2C}\right)S^{-1}-\frac{\alpha}{6}.
\end{equation}
 The former characterizes the natural precession frequency of the
 spin.  The latter is the part of the Hamiltonian composed of variable
 terms:
\begin{equation}\label{eqn:H2}
\mathcal H'=-g(\Shat\bdot\Jhat)-\frac{\alpha}{2}(\Shat\bdot\hhat)^2.
\end{equation}
Viewed in terms of the projection $\Shat=(X,Y,Z)$ onto the rotating
axes, equation (\ref{eqn:H2}) describes a family of parabolic cylinders opening
in the $+Y$ direction \citep{Colombo66}.  Solution trajectories for
$\Shat$ are curves of intersection between the parabolic cylinders and
the unit sphere.  The canonical equations of motion
$\dot\phi=\partial\mathcal H/\partial(S\cos\theta)$ and
$d(S\cos\theta)/dt=-\partial\mathcal H/\partial\phi$ become
\begin{equation}\label{EOM}
\frac{d\Shat}{dt}=g(\Shat\cross\Jhat)+\alpha(\Shat\bdot\hhat)(\Shat\cross\hhat).
\end{equation}
The equilibrium solutions of (\ref{EOM}) are Cassini states.
Constant obliquity requires that the spin axis be coplanar with
$\Jhat$ and the orbit normal (i.e. $\phi=0$ or $\pi$), because
otherwise $\Shat\cross\Jhat$ and $\Shat\cross\hhat$ point in different
directions.  This statement is Cassini's third law.  Constant
$\phi$ requires that the equilibrium obliquity $\theta$ satisfy a
relation
\begin{equation}\label{Cassini}
0=g\sin(\theta- I)+\alpha\cos\theta\sin\theta
\end{equation}
that has either two or four roots for given values of $g/\alpha$ and
$I$.  By convention $\theta$ is negative when $\phi=0$ and positive
when $\phi=\pi$.  Equation~(\ref{Cassini}) is identical to that
obtained by finding the stationary points of $\mathcal H'$ on the unit
sphere, which are points of tangency between the unit sphere and
parabolic cylinders.  

Figure~\ref{fig:obliq} illustrates the equilibria admitted by
equation~(\ref{Cassini}) for $I=0.1$ (left panel) and $I=1.0$ (right
panel).  In the limit that the orbit precesses slowly ($|g/\alpha|\ll
1$) the term derived from $V$ dominates the Hamiltonian.  The mean
gravitational potential is minimized when the spin is aligned or
antialigned with the orbit normal (states 1 and 3, respectively), and
maximized when the spin tips into the plane of the orbit, pointing in
the $+Y$ direction (state 2) or the $-Y$ direction (state 4).  States
1, 2, and 3 are stable against small displacements, which result in
libration around the equilibrium point.  The torque $\mathbf
S\cross\mathbf g$ is responsible for libration within a domain around
state 2, which would otherwise not be stable.  State 4 is a saddle
point of $\mathcal H'$, lying on the trajectory that separates the
domains of the other three states (i.e. the separatrix), and is
therefore unstable.  As $|g/\alpha|$ increases the domain of state 2
expands while that of state 1 contracts, until a critical point is
reached at which states 1 and 4 merge and vanish.  In the limit of a
rapidly precessing orbit ($|g/\alpha|\gg 1$) the planet responds to
the time-average position of the orbit normal, namely $\Jhat$.
$\mathcal H'$ is then maximized for spin aligned with $\Jhat$ (state
2) and minimized for spin antialigned with $\Jhat$ (state 3).
Comparison of the left and right panels illustrates that, at larger
inclinations, the domain around state 2 is broader because it lies
closer to $\Jhat$.

\citet{Winn_Holman05} pointed out that Cassini state 2 is the most
favorable configuration wherein a planet could maintain a significant
obliquity.  The obliquity of state 1 is large only for a coincidental
similarity of $g$ and $\alpha$ (near the critical point), which are
unrelated parameters.  As for state 3, we show in
\S\ref{sec:equilibration} that no such equilibrium exists in the
presence of a dissipative tidal torque.

\section{Tidal interactions}\label{sec:tides}

As we have just seen, a torque must be exerted on the planetary orbit
to keep it in Cassini state 2.  This torque could be tidal, involving
the oblateness and obliquity of the central star, or it could be due
to a third body in an inclined orbit.  We examine both possibilities
below.  Tidal dissipation must occur in the planet at some level if it
remains in Cassini state 2.  We compare the relative importance of
dissipation in the planet and in the star for the evolution of the
orbital energy and angular momentum.

\subsection{Orbit precessed by stellar oblateness} \label{subsec:oblateness}

Given dissipation in a body, the spin will move towards alignment with
the orbit normal.  However, given reasonable values for the star's
energy dissipation rate, the tide of a hot Jupiter is too weak to
cause the angle between the star's spin and its orbit to change
substantially (the time scale for such change is similar to equation
(\ref{eqn:tsync}) below).  Therefore, measurements of alignment among
extrasolar planets (\citealt{Queloz_etal00}; \citealt{Winn_etal05};
\citealt{Winn_etal06b}) are understood to be faithful indicators of
the primordial inclination.  A system that is not perfectly aligned
will undergo precession analogous to that of the Earth's spin as the
Earth's rotational bulge is torqued by both the moon and the sun.  As
an extrasolar planet causes its host star to precess, the planet's
orbit also precesses, conserving total angular momentum.  By this
mechanism, the orbits of hot Jupiters may precess at rates that are fast enough to
be observable by transit timing \citep{2002M}.  In such systems, the
star's spin angular momentum ${\bf S_{\star}} = C_{\star} {\bf
\Omega_{\star}}$ has a magnitude comparable to, but somewhat smaller
than, that of the orbital angular momentum ${\bf L}$, and the planet's
spin angular momentum ${\bf S_p} = C_{p} {\bf\Omega_p}$ has negligible
magnitude.  The total angular momentum ${\bf J} = {\bf L} + {\bf
S_{\star}} + {\bf S_{p}} $ is located between the stellar spin and
orbit normal, but it lies slightly outside of the plane defined by
${\bf L}$ and ${\bf S_{\star}}$ unless ${\bf S_{p}} $ lies in that
plane.  If the time average of ${\bf S_{p}}$ is kept out of that
plane, then the misalignment between stellar spin and the orbit normal
can evolve, even without dissipation in the star (see
\S\ref{sec:cassini_evolve}).

\subsection{Orbit precessed by third body} \label{subsec:thirdbody}

Another source of orbital precession is a third body.  There are two
known transiting planets in binaries (HD~189733b \& HAT--P-1b), but the
stellar companions would cause planetary orbital precession on a
timescale much longer than the host star's stellar oblateness would,
so their effect is negligible.  No system with a transiting planet has
a known second planet.

Suppose there is an undetected second planet on an exterior orbit.  It
will perturb the eccentricity of the transiting planet on a short
timescale, which will be continually damped by solar tides, which
would also heat the planet.  However, as the transiting planet's orbit
shrinks due to this continual excitation and damping of eccentricity,
the rate $|g|$ at which it is precessed by the exterior planet will
decrease.  Also, as the orbit shrinks, $\alpha$ increases.  If the
planet started with $|g|/\alpha > (|g|/\alpha)_{\rm crit}$, Cassini
state 2 was the only equilibrium available.  As $|g|/\alpha$ slowly
drops, the spin axis will tip over as $|g|/\alpha$ crosses
$(|g|/\alpha)_{\rm crit}$ and the spin remains in state 2 (see
Fig.~\ref{fig:obliq}).  Therefore a mixture of eccentricity and
obliquity heating is possible, if Cassini states are stable despite
dissipation.

The amount of angular momentum that the second planet can absorb from the
first as its orbit shrinks can constrain the parameters of the second
planet.  Let $L$ and $L_2$ be the orbital angular momentum of the
transiting planet and the second planet, respectively.  Suppose that
$L$ and $L_2$ are the dominant angular momenta of the problem such
that, by the law of cosines,

\begin{equation}
J^2 = L^2 + L_2^2 + 2 L L_2 \cos i_{\rm rel}, \label{eqn:JL2}
\end{equation}
where $i_{\rm rel}$ is the relative inclination of the two planets'
orbits.  The semimajor axis and eccentricity of the second planet are
both conserved on secular timescales if the semimajor axis ratio is
large, which can be shown by taking the potential of the second planet
to quadrupole order and integrating over the orbits of both planets
\citep{2000FKR}.  Therefore, the only way the orbit of a second planet
can absorb angular momentum from the dissipating planet is by
reorienting.  As the planets approach coplanarity, the orbital precession
period becomes too long for the dissipating planet to remain in
Cassini state 2.  The obliquity damps to zero and loss of energy and
angular momentum halts.  For the final condition of $i_{\rm rel}=0$,
we therefore have $J = L_f + L_2$, where $L_f$ is the final angular
momentum of the dissipating planet.  Constancy of $J$ gives:
\begin{equation}
L_f = (L_i^2 + L_2^2 + 2 L_i L_2 \cos i_{\rm rel,i} )^{1/2}- L_2 , \label{eqn:Lfinal}
\end{equation}
where $L_i$ and $i_{\rm rel,i}$ are the initial angular momentum of
the dissipating planet and the initial relative inclination,
respectively.  Solving for the angular momentum of the second planet
gives
\begin{equation}
L_2 = \frac12 \frac{L_i^2 - L_f^2}{L_f-L_i\cos i_{\rm rel,i}}. \label{eqn:L2}
\end{equation}
For any solution to exist, we must have $i_{\rm rel,i} > i_{\rm rel,crit} =
\cos^{-1}(L_f/L_i)$.  This critical value corresponds to the case
in which $L_2 \gg L$ and essentially all of the reorienting is done
by $L$.  

If we suppose that a particular planet is currently in Cassini state
2, with its semimajor axis still shrinking, then $L_{\rm now} > L_f$.
If a second planet is the cause of orbital precession, its angular momentum
must be greater than $L_2$ of equation~(\ref{eqn:L2}), taking $L_f
\rightarrow L_{\rm now}$.  This constraint may be very restrictive,
for a presumed value of $i_{\rm rel,i}$, if we can estimate $L_i$.
Tidal dissipation is a strong function of semi-major axis; from the
equations of \S\ref{sec:numerical} one may derive an approximate
relation $\dot{L} \propto - L^{-12}$.  If the rate of angular momentum
evolution $\dot{L}$ is observed or is theoretically inferred (for
instance, by an inflated radius), this relation can be integrated back
in time for the age of the system to obtain $L_i$.  Auxiliary
theoretical arguments may constrain $i_{\rm rel,i}$, so the angular
momentum $L_2$ of an undetected second planet may be given a lower
bound, if it is forcing obliquity tides.  The example of HD~209458b
will be examined in \S\ref{sec:discussion}.

\subsection{Dissipation} \label{subsec:tidalQs}

The nature of tidal dissipation in Jovian planets and late-type
main-sequence stars remains a subject of research.  Such bodies are
largely fluid, and the tidal Reynolds number based on the true
microscopic viscosity is enormous.  Therefore, most theories for the
dissipation invoke resonant coupling to stellar convection
\citep{Zahn66b}, or to modes of oscillation
\citep{Zahn70,Ogilvie_Lin04,Wu05b}.  Where these mechanisms can be
worked out with confidence, they often fall short of explaining
observationally inferred rates of circularization or synchronization.
Other mechanisms show promise but require difficult and uncertain
calculations.  Therefore, in the present work, we rely upon empirical
scalings from observations.  Following \citet[henceforth GS66]{GS66},
we parametrize the dissipation by a tidal quality factor $\tidalQ$
defined as the ratio of the maximum stored energy in the tidal
distortion of the star to the energy dissipated per radian of the
tidal cycle.  The ``equilibrium tide'' approximation is used, as if
the tide were static, so the stored energy is potential.  In this
formulation, the phase lag between the applied tidal potential and the
response is $(2\tidalQ)^{-1}$.  The torque on a nonrotating body of
mass $M_1$ and radius $R_1$ due to a companion of mass $M_2$ at
separation $r$ is
\begin{equation}
  \label{eq:GStorque}
  \Gamma_1 = \frac{3k_1}{\tidalQ_1}\frac{GM_2^2 R_1^5}{r^6}\,.
\end{equation}
The quantity $k_1$ is the apsidal-motion constant (which is half the
Love number), which GS66 effectively replace by $3/4$ as appropriate
for a body of uniform density---a reasonable approximation for a solid
planet or moon.  The structures of solar-type stars and Jovian planets
are better approximated by polytropes of indices $n_\star=3$ and
$n_p=1$, respectively, for which $(k_\star,k_p)\approx(0.014,0.26)$.
In all tidal dissipation rates, however, $\tidalQ$ and $k$ appear in
the same combination that one sees in equation~(\ref{eq:GStorque}) above,
so it is convenient to frame the discussion in terms of an effective
quality factor\footnote{In GS66's original notation, $\tidalQ$ and
$\tidalQp$ differ by a term involving elasticity, which can be
significant for solid planets but is presumed to be negligible in our
case.  Our $\tidalQp$ reduces to their $\tidalQ$ for a uniform-density
body.}  $\tidalQp\equiv \tidalQ\times 3/(4k)$.

Values for the quality factor can be inferred from orbital properties
of appropriate binary stars, extrasolar planets, and satellites of
Solar System planets.  Since it is simply a parametrization of
ill-understood dissipative processes, $\tidalQp$ may depend upon tidal
period, and also on subtle details of internal structure.  As an
example of the latter, \citet{Ogilvie_Lin04} state that their
inertial-wave mechanism depends upon the existence of a rocky
planetary core; there is evidence from the measured radii of
transiting planets that their cores vary considerably in size and
mass.

With these caveats in mind, we review some of the observational
inferences regarding $\tidalQp$.  GS66 estimated that the mean-motion
resonances of Io, Europa, and Ganymede require Jupiter's quality
factor to lie in the range $10^5\lesssim\tidalQp_J\lesssim10^6$,
unless that resonance is accidental or short-lived.  The relevant
tidal period in this case is $6.5$~hr, one half the synodic period
between Jupiter's rotation and Io's revolution.  For spectroscopic
binary stars, tidal circularization is easier to recognize than tidal
synchronization, since orbital eccentricity is measurable from the
radial-velocity curve.  \citet{Mathieu_etal92} and \citet[and
references therein]{Mathieu_etal04} find that the circularization
period ($P_{\rm circ}$) separating circular from measurably eccentric
binary orbits in stellar clusters increases with cluster age
beyond $1$~Gyr, indicating that significant tidal
dissipation occurs on the main sequence.  For M67 and NGC~188, for
example, $P_{\rm circ}=12.4\,{\rm d}$ and $P_{\rm circ}=15\,{\rm d}$
are quoted, with corresponding cluster ages of $5$ and $7$~Gyr,
respectively, whereas clusters younger than $1$~Gyr are said to be
consistent with $P_{\rm circ}\approx 8$~d, perhaps reflecting rapid
circularization during protostellar phases \citep{Zahn_Bouchet89}.  
To relate such observations to tidal dissipation rates, it
is conventional to suppose that $P_{\rm circ}$ is the period at which
eccentricity has decayed by three e-foldings.
Adapting GS66's
formula for the damping rate due to star~1 alone, we have
\begin{equation}
  \label{eq:GSdedt}
  \left(\frac{d\ln e}{dt}\right)_1= -\frac{171}{16\tidalQp}\frac{M_2}{M_1}
\left(\frac{R_1}{a}\right)^{5}~n,
\end{equation}
in which $a$ is the binary semimajor axis, $n=(GM_{\rm
tot}/a^3)^{1/2}$ is the mean motion of the orbit, and synchronous
rotation has been assumed.  Applying this
to a binary consisting of two sunlike stars and summing the dissipation
rates in both stars, we estimate from
M67 that $\tidalQp_\odot\approx 3.5\times 10^5$, and from NGC~188 that
$\tidalQp_\odot\approx 2.2\times 10^5$.

The fact that the orbits of most extrasolar planets with periods less
than $\sim 5$ days and masses $\gtrsim M_J$ are essentially circular
provides another constraint on $\tidalQp_p$, if we assume that this is
due to tidal dissipation in the planet.\footnote{
At the time of writing, based on the catalog at
{\tt\url{http://vo.obspm.fr/exoplanetes/encyclo}}, we find $P_{\rm
circ}\gtrsim 5\,{\rm d}$.  GJ 436b is the only planet with
$P<5\,{\rm d}$ that has an eccentricity securely above $0.1$, which
\cite{2007M} attribute to either ongoing eccentricity pumping or an
unusually high $\tidalQ$ for this rather low mass ($M \sin i \approx
23 M_\earth$) planet.}  Again taking $|d\ln e/dt|^{-1}=t/3$ at
$P_{\rm circ}$ and using $t= 5~\mbox{Gyr}$ as a representative age,
one has
\begin{equation*}
\tidalQp_p\approx 3\times 10^5\left(\frac{P_{\rm circ}}{5\,{\rm d}}\right)^{-13/3}
\left(\frac{M_\star}{M_\odot}\right)^{-2/3}\left(\frac{M_J}{M_p}\right)
\left(\frac{R_p}{R_J}\right)^5\,.
\end{equation*}
Alternatively, circularization might be due to dissipation in the
star.  However, equation~(\ref{eq:GSdedt}) predicts that $d\ln e/dt$ is
dominated by whichever body has the smaller value of $\tidalQp
M^2/R^5$.  This is probably the planet, as discussed below.

It is reassuring that these various estimates for $\tidalQp$ in
Jupiter, extrasolar planets, and sunlike stars are comparable; in
fact, one might boldly conclude from the evidence above that there
is a universal value $\tidalQp\sim 3\times 10^5$.
Many theories of tidal dissipation, however, predict that
$\tidalQp$ should scale with tidal period.  For equilibrium tides
subject to a frequency-independent effective viscosity (perhaps
associated with turbulent convection), it can be shown that
$\tidalQp\propto P$.  With this scaling, the values estimated for
Jupiter from its Galilean moons should be scaled up by an order of
magnitude for extrasolar planets in $\sim4$~d orbits.
\citet{Goldreich_Nicholson77} argued that when the tidal period
is less than the turnover time of the energy-bearing convective
eddies, the effective viscosity they provide should be suppressed by a
factor $\sim(P_{\rm tide}/T_{\rm eddy})^2$; this would lead to a
quality factor \emph{decreasing} with increasing period as
$\tidalQp\propto P^{-1}$.  \citet{Goldman_Mazeh91} suggested that
this scaling fits the observed trend of binary circularization periods
with age better than an increasing or constant $\tidalQp$.  However,
if the scaling $\tidalQp\propto P^{-1}$
results from a suppression of convective dissipation at
short periods, then it is very unlikely that convection can account
for the inferred $\tidalQp$ of Jupiter at all, as
\citet{Goldreich_Nicholson77} point out.  The inertial-wave mechanism
of \citet{Ogilvie_Lin04}, on the other hand, apparently predicts a
quality factor that is independent of tidal frequency for a fixed
ratio between the tidal and rotation frequencies.

In the case of hot Jupiters, dissipation in the planet is likely to be
more important than dissipation in the star.  On the one hand, it
follows from equation~(\ref{eq:GStorque}) that the ratio of the
synchronization torques on the spins of the two bodies is
\begin{equation}
  \label{eq:torqueratio}
  \frac{\Gamma_\star}{\Gamma_p}= \frac{\tidalQp_p}{\tidalQp_\star}\,\times\,
\frac{\bar\rho_p^2 R_p}{\bar\rho_\star^2 R_\star}\,,
\end{equation}
where $\bar\rho_i\equiv 3M_i/4\pi R_i^3$ is the average density of
body $i$.  The same ratio determines the relative importance of the
two bodies for eccentricity damping.  The densities of the star and
planet are typically comparable, and the mass ratio $M_p:M_\star\sim
10^{-3}$, so the second factor on the right hand side above is
$\lesssim (M_p/M_\star)^{1/3}\sim 0.1$.  So the torque on the planet
is greater unless the star has a substantially smaller $\tidalQp$.  On
the other hand, the synchronization timescale due to the torque on the
star is approximately
\begin{equation}
  \label{eqn:tsync}
  t_{\rm sync,\star}\equiv \frac{(M_p a^2 +C_\star)}{\Gamma_\star}
\left|n\hhat-\b{\Omega}_\star\right|~
\approx 15\,\frac{\tidalQp_\star}{3\times 10^5}\left(\frac{P}{4\,\rm d}\right)^{13/3}
\left(\frac{\bar\rho_\star}{\bar\rho_\odot}\right)^{5/3}\left(\frac{M_J}{M_p}\right)
\,{\rm Gyr}\,,
\end{equation}
in which $C_\star\approx 0.06M_\star R_\star^2$ is the stellar moment
of inertia.  Note that the stellar and orbital angular velocities are
differenced as vectors, to allow for possible stellar obliquity.  In
the final numerical form, we have neglected $\b{\Omega}_\star$
compared to $n\hhat$ and taken $C_\star\approx M_p a^2$: in fact
$C_\star/(M_p a^2)\approx 0.6 (P/{\rm 4\,d})^{-4/3}$ for a sunlike
star and a Jupiter-mass planet.  Thus, the stellar spin will not have
been substantially altered by its tidal interaction with the planet
unless $\tidalQp_\star < 10^5$, contrary to the evidence of binary
stars.  Of course, $\tidalQp_\star$ may depend upon the companion.
Theory suggests that $Q_\star$ may be \emph{greater} under the tidal
influence of a planetary rather than a stellar companion
\citep{OL2007}.

Moreover, the spectroscopic $(v\sin i)_\star$ values of exoplanetary
hosts indicate that the stars rotate subsynchronously.  To emphasize
this point, in Figure~\ref{fig:skumanich} we plot the values of $(v\sin i)_\star$, relative to the synchronous value, for the known transiting
planets.  The value of $\sin i$ should be very close to $1$ if there
is approximate alignment between stellar spin and planet orbit, as has
been observed in four cases \citep[and references therein]{Winn06}.
All of the stars rotate much slower than their planet's orbit.  In
fact, their spin rates are largely consistent with a standard
Skumanich law---an empirical relation $v\propto t^{-0.5}$ that
describes the braking with age $t$ of late-type dwarf stars due to the
stellar wind \citep{1972S}.  A suitable reference for calibrating the
Skumanich law is \citet{Valenti_Fischer05}'s spectroscopic catalogue
of 1040 nearby F, G, and K stars that have been targeted by planet
searches.  We calibrate separately for six mass bins, the first with
$0.7\leq M_\star/M_\odot<0.9$, and the remaining five bins spaced
between $0.9-1.4\msun$, each with width $0.1\msun$.  In all bins a
better fit to the data can be obtained with a power law steeper than
$-0.5$, but we ignore this systematic and depict standard Skumanich
laws for simplicity.  Scatter around the fits is $\approx 10\%$.  Only
HD~209458 appears to be rotating more rapidly than would be predicted
by magnetic braking alone (and only marginally so, given the scatter
and systematic uncertainties).  The tidal torque these stars receive
from their planets is apparently weak compared to the torque from
their magnetospheres.  Stars HAT--P-1 and TrES-1 are the relatively
slow rotators of the group; however, note that on this figure, the Sun
would also be rotating slowly by a factor of 2.

We conclude that for the purposes of the present paper, tidal
dissipation in the host star can be neglected.

\section{Spin and orbit equations of motion}\label{sec:numerical}

Here we introduce the equations of \cite{2001EK} for the spin and
orbit evolution of two fluid bodies.  The equations of motion are for
the orbital elements in the secular approximation.  That is, the
Newtonian equations of motion are integrated over one orbit, and what
remains is time derivatives for the orbital elements.  This
formulation allows us to perform integrations quickly that represent
dynamical evolution over billions of years.  The parameters of energy
dissipation in this model are based on the assumption that the energy
loss rate is proportional to the time rate of change of the quadrupole
tensor \citep{1998EKH}.  A prolate bulge is raised on each body by the
tidal potential of its companion.  If the rotation is not
synchronized, energy dissipation drags the bulge away from the
instantaneous direction of the companion.  The familiar consequences
are that the spins of the bodies synchronize with the orbit and that
the orbit's eccentricity is damped as its energy is sapped at nearly
constant angular momentum.  Another effect quantified by this model is
the spins coming into alignment with the orbit normal, which is the
key piece of technology we need to investigate the relationship of
tidal synchronization to Cassini states.

We shall specialize to the limit of zero orbital eccentricity, for a
few reasons.  First, the eccentricity will damp to zero on a timescale
short compared to the age of the system.  Second, the effect of
eccentricity damping in hot Jupiters has been pursued by other authors
\citep{Bodenheimer_etal01,Levrard_etal07}, so this approximation more
clearly focuses attention on the contribution of obliquity to tidal
dissipation.  Finally, this approximation considerably simplifies the
equations of \cite{2001EK}; for instance, the only included effect of
general relativity is to cause the apse to precess, which has no
physical effect when $e=0$.

Let the two masses equal $M_\star$ and $M_p$, of total mass $M$, and
of reduced mass $\mu = M_\star M_p / M$.  Let ${\bf h}$ denote the
orbital angular momentum per unit reduced mass, of magnitude $h$ and
direction $\hhat$, and let $\ehat$ and $\qhat$ be reference directions
in the plane of the orbit such that $\ehat\cross\qhat=\hhat$.  Let the
vector spins of the two bodies be ${\bf \Omega_\star}$ and ${\bf
\Omega_p}$ with rotational moments of inertia $C_\star$ and $C_p$.
Therefore, with our specialization to $e=0$, there are $9$ dynamical
variables: the $9$ components of the vectors ${\bf h}$, ${\bf
\Omega_\star}$, and ${\bf \Omega_p}$.  The equations conserve the
total angular momentum.  The equations of motion are:

\begin{eqnarray}
\frac{1}{h} \frac{d {\bf h}}{dt} &=& (\mathcal Y_\star + \mathcal Y_p)
\ehat - (\mathcal X_\star + \mathcal X_p) \qhat -
(\mathcal W_\star + \mathcal W_p) \hhat \label{eqn:ekehdot} \\
C_i \frac{ d {\bf \Omega_i} }{ dt } &=& \mu h ( - \mathcal Y_i
\ehat + \mathcal X_i \qhat + \mathcal W_i \hhat ),
\label{eqn:ekeomegadot}
\end{eqnarray}
where 
\begin{eqnarray}
\mathcal X_i &=& - k_i \frac{M}{M_i} \frac{1}{ n} \Big( \frac{R_i}{a} \Big)^5
\Omega_{ih} \Omega_{ie} - \frac{\Omega_{iq}}{2  n t_{Fi}}\label{eqn:ekeX} \\  
\mathcal Y_i &=& - k_i\frac{M}{M_i} \frac{1}{ n} \Big( \frac{R_i}{a} \Big)^5
 \Omega_{ih} \Omega_{iq} + \frac{\Omega_{ie}}{2  n t_{Fi}} \label{eqn:ekeY} \\ 
\mathcal W_i &=& \frac{1}{t_{Fi}} \Big(1 - \frac{\Omega_{ih} }{  n} \Big),
\label{eqn:ekeW} 
\end{eqnarray}
and $i$ is either $p$ or $\star$.  $R_i$ is the radius of each
body, $a = h^2/(GM)$ is the semi-major axis, $ n = (GM)^2 h^{-3}$ is
the mean motion, $k_i$ is the apsidal motion constant,
and $t_{Fi}$ is a tidal friction timescale:
\begin{equation}\label{eqn:tf}
t_{Fi} = \frac{4\tidalQp_i}{9}\left(\frac{a}{R_i}\right)^5 
\frac{M_i^2} {\mu M}n^{-1}.
\end{equation}
The integrations in \S\ref{sec:cassini_evolve} hold the product $\tidalQp_i n^{-1}$
constant during the evolution, as recommended by the equilibrium tide
model of \citet{1998EKH}.

\subsection{Equilibration}\label{sec:equilibration}

In equations (\ref{eqn:ekeX}) and (\ref{eqn:ekeY}), the first term on
the right hand side is a non-dissipative torque that gives rise to
nodal precession of the spins and orbit.  The second term in these
equations and the term on the right hand side of equation
(\ref{eqn:ekeW}) are due to tidal dissipation.  Since dissipation in
the star is negligible, we may set $t_{F\star} \rightarrow \infty$,
and specialize the discussion to dissipative torques on the planet.

In equilibrium the magnitude $\Omega_p$ of the planetary spin angular
velocity is constant.  Therefore
\begin{equation} \label{eq:zerot}
\left\langle\b{\Gamma\cdot\Omega}_p\right\rangle=0,
\end{equation}
where $\b{\Gamma}$ is the total torque on the planetary spin, and the
angular brackets denote an average over the orbital mean motion.  It
suffices to replace $\b{\Gamma}$ by the tidal (\emph{i.e.}
dissipative) component, since the precessional (\emph{i.e.}
nondissipative) part already satisfies (\ref{eq:zerot}).  In the
equilibrium tide, the \emph{instantaneous} tidal torque takes the form
\begin{equation} \label{eq:inst}
\b{\Gamma}_t = -\kappa \b{\hat r\times}(\b{\hat r\times u}),
\end{equation}
where $\b{u} \equiv n \b{\hat{h}}-\b{\Omega}_p$ and $\b{\hat r}$ is a
unit vector from the star toward the planet.  The logic behind
equation~(\ref{eq:inst}) is that if the tidal bulge is a prolate distortion
with major axis direction $\b{\hat b}$, then $\b{\Gamma}_t \propto
\b{\hat b\times\hat r}$. Absent dissipation, $\b{\hat b}=\b{\hat r}$
so that the torque vanishes.  Dissipation gives the bulge a phase lag
of magnitude $\approx (2Q)^{-1}$ with a direction determined by the
angular velocity of the tide seen in a frame spinning with the planet.
This angular velocity is $\b{u}$, so that $\b{\hat b} \approx \b{\hat
r}+(2Q)^{-1}(\rhat\cross{\bf u})/|\rhat\cross{\bf u}|$.  Comparison
with the equations of motion shows that $\kappa=\mu h(nt_{Fp})^{-1}$.
Substituting equation (\ref{eq:inst}) into equation (\ref{eq:zerot})
gives
\begin{equation} \label{eq:subst}
\b{\Omega}_p\b{\cdot}(\b{u}-\b{u\cdot}\langle\b{\hat r\hat r}\rangle)=0,
\end{equation}
since the precession of $\b{\Omega}_p$ \& $\b{u}$ can be neglected on
orbital timescales.  Note that $\kappa$, which measures the strength
of the tide, drops out.  Now it is easy to show that\footnote{By
symmetry, $\langle\b{\hat r\hat r}\rangle$ annihilates $\b{\hat h}$
and is a multiple of the identity $\mathbf{I}$ for any vector lying in
the orbital plane.  The factor $(1/2)$ follows by considering the
orbit average of $(\b{d\cdot\hat r})^2$ for any constant vector
$\b{d}$ in the plane.} $\langle\b{\hat r\hat
r}\rangle=\onehalf(\mathbf{I}-\b{\hat h\hat h})$, which, in
combination with (\ref{eq:subst}), yields $\b{\Omega}_p\b{\cdot
u}+(\b{\Omega}_p\b{\cdot\hat h})(\b{u\cdot\hat h})=0$.  Solving for
the equilibrium spin frequency of the planet, we have the correct
statement of Cassini's first law:
\begin{equation}\label{eqn:omegap}
\Omega_p = \frac{2  n}{\cos \theta + \sec \theta}.
\end{equation}
This is the exact solution of the equations of motion when
$\dot\Omega_p=0$.  No equilibrium exists for $\theta>\pi/2$; in
particular, Cassini state 3 becomes unstable in the presence of the
dissipative torque.  Generally the synchronization time is comparable
to the parallelization time (as we show next), so $\Omega_p
\rightarrow n$ as $\theta \rightarrow 0$, but if there is a mechanism
for maintaining non-zero obliquity, like trapping in Cassini state 2,
$\Omega_p < n$.  Equation (\ref{eqn:omegap}) can be understood as
pseudo-synchronization when obliquity is held at a constant value; an
oblique and tidally evolved planet can be expected to be a slow
rotator.  These points were made by \cite{Levrard_etal07}, who gave an
expression for $\Omega_p$ with non-zero eccentricity (the dependence
on $\theta$ is separable from the dependence on $e$).

Neglecting orbital precession, the dissipative torque from equations
(\ref{eqn:ekeomegadot})--(\ref{eqn:ekeY}) will cause parallelization;
the component of the spin perpendicular to the orbit
$(\Omega_{p\perp})$ decays approximately exponentially on a timescale
$\tau_{\perp} = 2 C_p n (\mu h)^{-1}\, t_{Fp}$.  The deviation from
synchronization $(\Omega_{ph}-n)$ also decays, but on a timescale
$\tau_{s} = \onehalf\tau_{\perp}$.  The twice longer timescale for
parallelization is because the parallelization torque takes on smaller
values in parts of the orbit, whereas the synchronization torque is
constant.  For precessing orbits that trap the spin in Cassini state
2, equation (\ref{eqn:omegap}) applies after a few parallelization
times $\tau_\perp$, although it has no dependence on the tidal
dissipation rate.

\subsection{Cassini states with dissipation} \label{sec:cassiniwdiss}

The Cassini state formalism may be generalized to include ongoing
tidal torques.  For the generalized Cassini state we require that all
the angular momenta vectors are stationary with respect to each other,
even as the whole system precesses uniformly about the ${\bf J}$ axis.
Cassini states are defined on timescales over which the orbital precession
frequency $g$ is nearly constant; on tidal dissipation timescales, $g$
varies and the state evolves.

In the presence of dissipation, ${\bf \Omega_p}$ is shifted from the
usual Cassini state; it does not lie in the plane defined by ${\bf h}$
and ${\bf \Omega_\star}$.  The reason is that dissipation introduces
terms that attempt to align the planet's spin with its orbit, which
would cause ${\bf \Omega_p}$ to slowly spiral in towards ${\bf h}$.
In particular, the second terms of equations~(\ref{eqn:ekeX}) and
(\ref{eqn:ekeY}) cause a damping of the $e$- and $q$-components of the
spin, according to equation~(\ref{eqn:ekeomegadot}).  The system is
able to avoid this drift by readjusting to a state in which the
planetary spin executes pure precession about ${\bf J}$.  If the
planet's spin and the orbit normal are stationary in a frame
precessing with uniform angular velocity ${\bf g}=g\Jhat$, then $ {\bf
\dot{\Omega}_p} = {\bf g}\cross{\bf \Omega_p}$ and ${\bf\dot h}={\bf
g}\cross{\bf h}$.  Three constraints are needed on the three
components of ${\bf\Omega_p}$.

To analyze the situation, let us align the reference direction ${\bf
\hat{q}}$ such that ${\bf J}$ lies in the ${\qhat-\hhat}$ plane,
making an angle $I$ with the orbit normal ${\bf \hat h}$ as in
Figure~\ref{fig:coords} (\emph{i.e.} ${\bf \hat{q}} = \Yhat$ and ${\bf
\hat{e}} = \Xhat$).  Anticipating that Cassini state 2 will be shifted
out of the ${\qhat-\hhat}$ plane, denote the phase shift angle
$\phi_s\equiv\pi - \phi$, as it is the supplement of $\phi$.  Let the
angle between ${\bf \Omega_p}$ and ${\bf h}$ be denoted $\theta$ and
the angle between ${\bf \Omega_\star}$ and ${\bf h}$ be denoted
$\theta_\star$.  Then, in the $eqh$ coordinate system, we have
\begin{eqnarray}
{\bf g} &=& g(0, \sin I, \cos I ), \label{eqn:j} \\ 
{\bf\Omega_p} &=&
\Omega_p(\sin \phi_s \sin \theta , \cos \phi_s \sin \theta , \cos
\theta), \label{eqn:omega} \\ 
{\bf\dot\Omega_p} &=& g\Omega_p(\cos
\phi_s \sin \theta \cos I - \cos \theta \sin I , -\sin \phi_s \sin
\theta \cos I , \sin \phi_s \sin \theta \sin
I). \label{eqn:omegadotvec}
\end{eqnarray} 
Note that $g<0$ because the nodes regress.
Equations~(\ref{eqn:ekeomegadot}), (\ref{eqn:ekeX}), and
(\ref{eqn:ekeY}) give the ratio between $({\bf\dot{\Omega}_{p}})_e$
and $({\bf\dot{\Omega}_{p}})_q$.  Equating that to the ratio given by
(\ref{eqn:omegadotvec}), we obtain the first constraint:
\begin{eqnarray}
\frac{-\cos\phi_s\cos\theta + \xi^{-1}\sin\phi_s }
{\sin\phi_s\cos\theta +\xi^{-1}\cos\phi_s } &=& 
\frac{-\cos\phi_s\sin\theta\cos I + \cos\theta\sin I}{\sin\phi_s\sin\theta\cos I}\,,
\nonumber\\[1ex]
\mbox{or equivalently,}\qquad 
\sin(\delta + \phi_s)/\sin(\delta) &=& \tan \theta \cot I, \label{eqn:phic} 
\end{eqnarray}
where
\begin{equation}\label{eqn:xival}
\xi \equiv 2 k_p\Omega_p t_{Fp} \frac{M}{M_p}\left( \frac{R_p}{a}\right)^5
= \frac{2}{3}Q_p\,\frac{\Omega_p}{n}\frac{M_p}{\mu}\quad \mbox{and} \quad \tan{\delta} \equiv (\xi \cos \theta)^{-1}\,. \nonumber
\end{equation}
Note that $\xi$ has been written in terms of $Q_p$ rather than
$Q'_p=3Q_p/(4k_p)$; nevertheless, it is likely to be quite large.

The second constraint, similarly obtained from the ratio between
$({\bf\dot{\Omega}_p})_h$ and $({\bf\dot{\Omega}_p})_q$, gives an
expression for the equilibrium spin rate that we have already derived
in equation~(\ref{eqn:omegap}).  The final constraint can be derived
by setting equal the precession rates of the planet's orbit and spin.
The former (from Eq.~\ref{eqn:ekehdot}) is essentially due to the
stellar oblateness since the rotational angular momentum of the planet
is negligible ($\mathcal X_p \ll \mathcal X_\star$, etc.):
\begin{eqnarray}
g &=& -\frac{1}{\sin I} \frac{| {\bf \dot{ h}} |}{|{\bf h}|} \nonumber\\
  &=& -k_\star\frac{M}{M_\star} \Big(\frac{R_\star}{a}\Big)^5 \,
        \frac{\Omega_\star^2}{ n}\frac{\sin \theta_\star \cos
        \theta_\star}{\sin I}.\label{eqn:g}
\end{eqnarray}
The planet's spin precesses about the total angular momentum,
maintaining with it an angle $\beta\equiv\cos^{-1}(\Jhat\bdot\Ophat)$,
at the rate
\begin{eqnarray}
g  &=& -\frac{1}{\sin \beta} \frac{| {\bf \dot{\Omega}_p} |}{|{\bf\Omega_p}|}\nonumber\\ 
   &=& -\frac{\mu h}{C_p}\frac{\sin\theta}{\sin\beta} \left\{ \Big[\frac{M}{M_p}
        \Big(\frac{R_p}{a}\Big)^5 \frac{2k_p}{1 + \sec^2 \theta} \Big]^2
        + \Big[ \frac{1}{ 2 n t_{Fp} \cos \theta } \Big]^2
        \right\}^{1/2}.\label{eqn:gp}
\end{eqnarray}
Solutions of equations~(\ref{eqn:omegap}), (\ref{eqn:phic}),
(\ref{eqn:g}), and (\ref{eqn:gp}) are generalized Cassini states.

Given the complexity, solutions for $\phi_s$ and $\theta$ will be
found by making a few approximations.  To first order in $\phi_s$ and
$\delta$, which are small when dissipation is weak, (\ref{eqn:phic})
becomes
\begin{equation}
\phi_s \approx (\xi\cos\theta)^{-1}(\tan\theta\cot I -1).
\label{eqn:phisfo}
\end{equation}
Setting equal (\ref{eqn:g}) and (\ref{eqn:gp}) with the above
approximation gives the following constraint for $\theta$:
\begin{multline}
\frac{\sin \theta}{\sin (\theta- I)} \left\{ \left[ \frac{M}{M_p}
\left(\frac{R_p}{a}\right)^5 \frac{2k_p}{1+\sec^2 \theta} \right]^2 +
\left[ \frac{1}{2 n t_{Fp} \cos \theta} \right]^2 \right\}^{1/2} \\ =
\frac{C_p}{\mu h}k_\star \frac{M}{M_\star}
\left(\frac{R_\star}{a}\right)^5 \frac{\Omega_\star^2}{ n}\frac{\sin
\theta_\star \cos \theta_\star}{\sin I}.
\label{eqn:thetafo}
\end{multline}
In this final equation, the right hand side is determined by
parameters that are approximately constant over a planetary spin
precession period.  On the left hand side, the quantities that are
uncertain are the planet's apsidal-motion constant and its tidal
friction timescale.  If the former is too big, or the latter is too
small, no real solution for $\theta$ is possible, and there is no
equilibrium corresponding to Cassini state 2.  Tidal friction plays an
important role here because with $t_{Fp} \rightarrow \infty$, there is
always a solution for $\theta$.  Indeed, equation~(\ref{Cassini}) is
recovered in this limit\footnote{The spin precession constant $\alpha$
is related to the apsidal motion constant by
\begin{equation*}
\alpha=k_p\frac{\mu h}{C_p}\frac{M}{M_p}\left(\frac{R_p}{a}\right)^5\frac{\Omega_p}{n}.
\end{equation*}}.

Starting from the values for $\Omega_p$, $\phi_s$, and $\theta$ given
by equations~(\ref{eqn:omegap}), (\ref{eqn:phisfo}) and
(\ref{eqn:thetafo}), a Newton-Raphson solver can efficiently find any
solution to equations~(\ref{eqn:omegap}), (\ref{eqn:phic}),
(\ref{eqn:g}), and (\ref{eqn:gp}).  Once $\theta$ is found,
(\ref{eqn:omegap}) can be used to find the planetary spin rate.  Our
derivation gives an especially physical understanding for why the spin
cannot be truly synchronous: there needs to be a small component of
${\bf \dot{\Omega}_{p}}$ in the direction of the orbit normal so that
the spin performs precession about the total angular momentum rather
than the orbit normal.

\section{Evolution of Cassini states} \label{sec:cassini_evolve}

First we will show the integration of equations~(\ref{eqn:ekehdot})
and (\ref{eqn:ekeomegadot}) for an example that has an extremely
tilted Cassini state 2.  To find such a state we needed to take an
extreme value for $k_p = 5\times 10^{-5}$, which corresponds to a
polytrope $n=4.65$, instead of a more reasonable value such as $k_p =
0.26$ for $n=1$.  This choice lessens the planetary spin precession
torque and dissipative torque relative to the rather weak orbital
precession torque arising from the rotational bulge of the star.  We
also chose a quite large value of $Q_p' = 2.3 \times 10^7$.  The
planet started with a semimajor axis of $a = 0.05$ AU.  The stellar
spin began with a period of 10 d and made an angle $45^\circ$ with the
orbit normal.  The stellar apsidal motion constant was set to $k_\star
= 0.0144$, appropriate for an $n=3$ polytrope.  The masses were
$M_\star = M_\odot$ and $M_p = 10^{-3} M_\odot$, radii $R_\star = 1.0
R_\odot$ and $R_p = 0.1 R_\odot$, and normalized moments of inertia
$C_\star / (M_\star R_\star^2) = 0.08$ and $C_p / (M_p R_p^2) = 0.25$.

See Figure~\ref{fig:state2settle} for the evolution into Cassini
state~2.  At first the planetary spin axis librates with large
amplitude, tracing curves of nearly constant Hamiltonian
(Eq.~\ref{eqn:H2}).  After about $20$ Myr, dissipation has caused the
system to settle to a state of quasi-equilibrium (for comparison,
$\tau_{\perp} = 4.3$ Myr).  Panel~(a) shows the evolution of the
angles between the total angular momentum $\J$ and the orbit normal
$\mathbf{n}$, between $\J$ and the stellar spin axis
$\mathbf{\Omega_\star}$, and between $\J$ and the planetary spin axis
$\mathbf{\Omega_p}$.  Panel~(b) is a parametric plot (with time as the
suppressed parameter) of the projection of the planetary spin vector
${\bf \hat \Omega_p}$ onto the $XY$ plane of Figure~\ref{fig:coords}.
Panel (c) zooms in on the final state of panel (b), revealing that, in
the presence of ongoing tidal dissipation, the equilibrium Cassini
state~2 no longer lies on the $Y$-axis; it is shifted towards positive
$X$ values by the tidal torque.

The long-term consequence of such a state will be shown next.  We
started with the same initial conditions, except set $Q_p' = 4.6
\times 10^6$ so that the evolution would proceed faster.  An
integration for $1.5$ Gyr is shown in Figure~\ref{fig:cassinilong}.
Panels~(a) and (b) are the same as those of
Figure~\ref{fig:state2settle}.  The timescale of damping into the
Cassini state is a few Myr, smaller than the time axis resolution.
Cassini state~2 now lies very far from the $Y$-axis, and evolves even
farther from it.  The state ``breaks'' when $Y=0$, then librates
around, and damps into, Cassini state~1.  The physical cause is that
the tidal torque (the last terms on the right hand sides of
Eqs.~\ref{eqn:ekeX} and \ref{eqn:ekeY}, which are proportional to
$a^{-9/2}Q'^{-1}$) overcomes the orbital precession torque (the first terms
on the right hand sides of Eqs.~\ref{eqn:ekeX} and \ref{eqn:ekeY},
which are proportional to $a^{-3}$), as $a$ decreases due to energy
dissipation.  Panel (c) shows the time dependence of the phase shift
$\phi_s$ away from the $X$ axis (\emph{cf.} Fig.~\ref{fig:coords}).
Most of the time is spent in the evolving equilibrium of state 2, and
the brief libration epochs that are prominent in panel (b) are not
resolved in this plot.  Panel (d) shows the semi-major axis evolution:
dissipation in the planet converts orbital energy to heat energy.
Panel (e) shows the evolution of the various angular momenta of the
problem, projected along the total angular momentum axis.  The angular
momentum lost by the planetary orbit is absorbed by the stellar spin.
However, in the absence of extremely strong dissipation in the star,
the torque from the planet cannot cause the stellar spin period to
change (Eq.~\ref{eqn:tsync}).  Panel (f) shows a parametric plot of
the stellar spin vector broken into two components: those parallel to
and perpendicular to the planet's orbit.  It is clear that the star's
spin absorbs the angular momentum of the planetary orbit by
reorienting, not by changing magnitude.
 
The non-zero value of $\phi_s$ causes the stellar spin to come into
alignment with the planet's orbit, without any dissipation in the
star, by the following mechanism.  The spin angular momentum of the
planet is not coplanar with the other angular momenta, which causes
the total angular momentum $\J$ to lie slightly outside of the plane
defined by the dominant angular momenta $\mathbf{L}$ and
$\mathbf{S_\star}$.  The evolution equations cause $\mathbf{S_\star}$
to precess about $\mathbf{L}$ (there is no spin-spin coupling), but
this non-coplanarity directs a small component of that torque towards
${\bf J}$.  Similarly, the motion of the orbital angular momentum is
not quite the pure precession of equation~(\ref{eqn:g}) because the
small coupling of the orbit to the rotational bulge of the planet
($\mathcal X_p$ and $\mathcal Y_p$ of Eq.~\ref{eqn:ekehdot}) causes
a secular motion of the orbit normal towards ${\bf J}$.  Therefore the
stellar spin and planet orbit come into alignment with the total
angular momentum, and thus with each other.  Recall that the
ingredients for this alignment are a close planet whose spin is
trapped in Cassini state 2, whose tidal dissipation causes that state
to shift in phase.  In the example we have given, a significant
spin-orbit misalignment persists because the planet's spin left
Cassini state 2 before all of the available angular momentum from the
star was tapped.

In integrations like that of Figure~\ref{fig:state2settle} (in which
$\phi_s = 0.1106$), the first-order estimate of $\phi_s$
(Eq.~\ref{eqn:phisfo}) is quantitatively quite good, and its fidelity
for larger values can be seen in Figure~\ref{fig:cassinilong}(c).  It
is remarkable that the total amount of orbital energy dissipated is
many orders of magnitude beyond that which is stored in the planet's
spin; it is approximately the binding energy of the planet!  The
average dissipation rate indicated by the orbital migration of
Figure~\ref{fig:cassinilong}(d) is $\dot{E} \approx 6.2 \times
10^{43}$~erg~s$^{-1}$, enough to inflate the planet's radius (during
the first 1.1 Gyr after Cassini state 2 is established).  However, we
achieved this result by adopting an apsidal moment constant almost
four orders of magnitude smaller than Jupiter's, corresponding to a
very strongly centrally concentrated body, in order to reduce the spin
precession rate and dissipative tidal torque.  Such a value for $k_p$
would be much more strongly discrepant with theory than the inflated
radius itself.
  
To seek the range of structural parameters that allow Cassini state~2
to be an equilibrium, we may use $\phi_s = 1$ in
equation~(\ref{eqn:phisfo}) to approximately indicate when Cassini
state 2 will break due to dissipation in the planet.  This first order
estimate actually is serendipitously rather precise: recognizing from
Figure~\ref{fig:cassinilong} that Cassini state 2 actually breaks when
$\phi_s = \pi/2$, we may substitute this value into
equation~(\ref{eqn:phic}).  The result is that the right hand side of
(\ref{eqn:phisfo}) equals 1.  That is, the full phase shift
equation gives $\phi_s = \pi/2$ when the first order estimate gives
$\phi_s = 1$, which can be seen in Figure~\ref{fig:cassinilong}(c). 
  
What is the maximum energy dissipation rate a planet may experience
while in Cassini state~2, driven by stellar bulge orbital precession?
A closed equation for $\dot{E}_{\rm max}$ may be written by making a
few approximations (all of which, we have verified, are
well-justified).  For realistic planetary structure, $k_p = 0.26$,
which requires $\theta \approx \pi/2$.  In the evaluations below, we
will also use $k_\star = 0.0144$, $R_\star=R_\odot$,
$M_\star=M_\odot$, $C_\star / (M_\star R_\star^2) = 0.08$,
$\Omega_\star=2\pi/(20\rm{d})$, $\theta_\star=0.1$, $R_p=R_{J}$,
$M_p=M_{J}$, $C_p / (M_p R_p^2) = 0.25$, and $n=2\pi/(4\rm{d})$.
Expanding $g$ (Eq.~\ref{eqn:gp}) to first order in $\pi/2 - \theta$,
while neglecting the dissipative term and setting $\sin \beta
\rightarrow 1$, then setting the resulting expression equal to
equation (\ref{eqn:g}) gives
\begin{equation} \label{eqn:thetaapprox}
\theta \approx \pi/2 - \left( \frac{k_\star R_\star^5 }{2 k_p R_p^5 } \frac{a C_p \Omega_\star^2}{G M_\star^2} \frac{\sin \theta_\star \cos \theta_\star}{\sin I} \right)^{1/2}.
\end{equation}
Second, the solution will have an extreme value for $\phi_s$, which we
may identify by setting the first order estimate
(Eq.~\ref{eqn:phisfo}) to unity.  This allows us to identify the
$Q_p$ parameter for such a maximally dissipating planet:
\begin{equation}
Q_{p,\rm min} = \frac34 \frac{M_\star}{M} \frac{(\pi/2 - \theta)^{-3}}{\tan I}.
\end{equation}
For the nominal values above, this means $Q_p \geq 8 \times 10^8$ for
Cassini state 2 to be an equilibrium, where equality holds for a state
barely in equilibrium, with the planet dissipating at the maximal
value.  That the population of hot Jupiters is inferred to have $Q_p \approx
10^6$ (\S\ref{subsec:tidalQs}) implies that a planet must be unusually non-dissipative to remain in state 2.  Finally, the energy
dissipation $\dot{E}_{\rm max} = G^2 M_\star M_p M h^{-3} \dot{h}$ may be
written
\begin{eqnarray} 
\dot{E}_{\rm max} &=& (9/4) \tidalQp^{-1} G M_\star^2 R_p^5 a^{-6} n \nonumber\\ 
&=& \frac{2^{1/2} k_\star^{3/2}}{k_p^{1/2}} \frac{ M^{3/2}}{M_\star^2}
  \frac{R_\star^{15/2}}{R_p^{5/2}} \frac{C_p^{3/2}
  \Omega_\star^3}{a^6} \left( \frac{\sin^3\theta_\star
  \cos^3\theta_\star}{\sin I \cos^2I}
  \right)^{1/2},\label{eqn:edotmax}
\end{eqnarray}
which evaluates to $\dot{E}_{\rm max} \approx 5 \times
10^{23}$~erg~s$^{-1}$ given the above values for the parameters.  The
maximum possible rate of heating due to orbital precession-stabilized
obliquity tides is comparable to the luminosity of gravitational
contraction.  That is, if a certain transiting planet is in Cassini
state 2, then tidal dissipation is not strong enough to inflate that
planet's radius.  Equation~(\ref{eqn:edotmax}) has a strong dependence
on semimajor axis, but even a Jupiter mass planet at its Roche limit
cannot dissipate a structurally significant power via Cassini state 2,
even if it somehow managed to have $Q_p \gtrsim Q_{p,\rm min}$.

Another way to assess stability is to follow \cite{Levrard_etal07} in
requiring that the dissipative torque does not destroy $\theta$
stability (that a real solution exists for $\theta$ in
Eq.~\ref{eqn:thetafo}).  For the above parameters, this gives
$\dot{E}_{\rm max} \approx 6 \times 10^{24}$~erg~s$^{-1}$, a factor of
$\sim10$ less restrictive than our criterion that the first order
estimate of $\phi_s$ is less than 1.  We interpret this to mean that
Cassini equilibrium is generally lost in the $\phi$ direction (as in
Fig.~\ref{fig:cassinilong}[b]), not the $\theta$ direction.

\section{Discussion and Conclusions} \label{sec:discussion}

We have calculated the stability and evolution of Cassini state 2 with
large obliquity $\theta$ in the presence of tidal dissipation.  We
have verified that in principle, a spin state can couple the orbit to
internal energy such that many times the energy of the spin is
dissipated.  However, it cannot dissipate the energy rapid enough for
that heat to make a structural difference.  Here we apply the concepts
we have developed to the archetypal inflated planet, HD~209458b, and
find that neither the stellar bulge nor a second planet can be
responsible for a planet-inflating Cassini state.

According to \cite{Bodenheimer_etal03}, if the large radius of
  HD~209458b is due to internal heating, the inferred heating rate
  $\dot E$ is at least $4 \times 10^{26}$~erg~s$^{-1}$.  Can this
  $\dot E$ be continually supplied by obliquity tides over the age of
  the planet, without the Cassini state breaking?
  
  In \S\ref{sec:cassini_evolve} we evaluated the hypothesis that the
  stellar bulge causes the orbital precession, finding that the tidal torque
  causes Cassini state 2 to fall out of equilibrium.  When numbers
  appropriate to the HD 209458 system---including a plausible
  planetary apsidal-motion constant---are used in equation
  (\ref{eqn:edotmax}), the maximum possible heating rate is $3 \times
  10^{24}$~erg~s$^{-1}$, a factor of $\sim 100$ too small to explain
  the inflated radius.
    
  Independent of this dissipative torque constraint, we note that the
  usable angular momentum in the spin, as the star comes into
  alignment, is quite small given the observationally indicated
  stellar obliquity.  The angular momentum that the stellar spin can
  absorb from the orbit before complete alignment, using
  equation~(\ref{eqn:Lfinal}) with $L_2 \rightarrow S_\star$ and
  $i_{\rm rel,i} \rightarrow \theta_\star$, is
  \begin{equation}
  \Delta L = L_f - L= (L^2 + S_\star^2 + 2 L S_\star \cos \theta_\star )^{1/2} - S_\star - L \,.
  \end{equation}
  So for HD~209458, taking $\theta_\star = 0.1$, we have $\Delta L
  \approx -2.1 \times 10^{46}$~g~cm$^2$~s$^{-1}$.  The quoted $\dot E$
  implies $\dot L \approx -1.9 \times 10^{31}$~g~cm$^2$~s$^{-2}$.
  Then, if the angular momentum is lost at a steady rate, the system
  will become coplanar and the high obliquity state will end in only
  $\sim35$ Myr, which is $\lesssim1\%$ of the system's age (5 Gyr;
  \citealt{CodySasselov2002}).  Under this hypothesis we are seeing
  the system at a very special time, just before it becomes coplanar.
  
\begin{figure}
\plotone{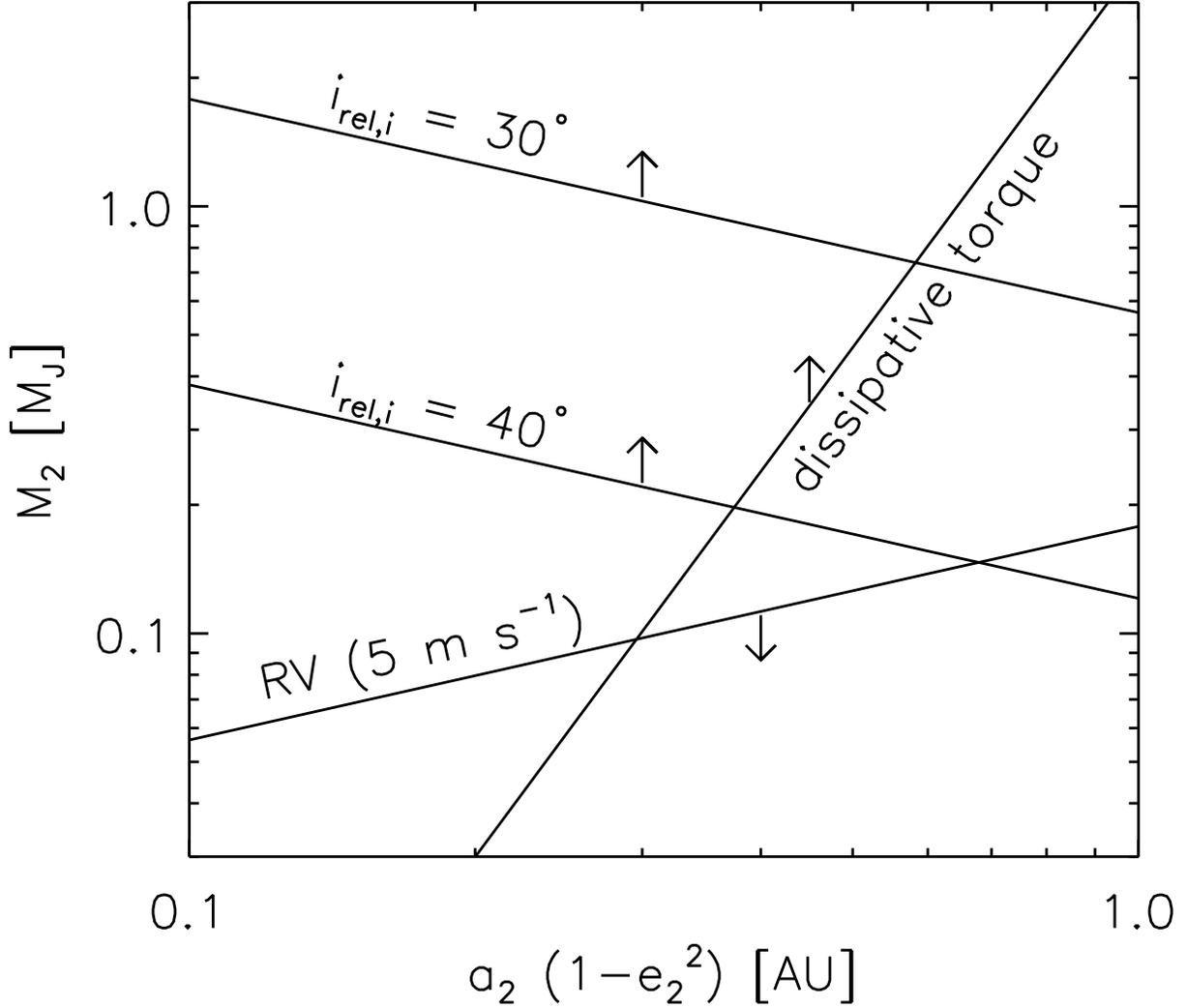}
\caption{ The limits of mass $M_2$ and orbital elements $a_2 (1-e_2^2)$ of a second planet for the hypothesis that it is responsible for the inflation
  of HD~209458b via a Cassini resonance.  Here $i_{\rm rel,i}$ is the
  initial relative inclination between the two planets, and the bounds
  so labeled reflect the tendency of the second planet's orbit to
  align as it absorbs the angular momentum shed by the orbit of
  HD~209458b ($L_2$ must be greater than that of eq. [\ref{eqn:L2}]).
  Values of $i_{\rm rel,i}>40^\circ$ are unlikely as the second planet
  would force Kozai eccentricity oscillations in the transiting
  planet.  The curve labeled ``dissipative torque'' is the lower limit
  of mass needed to overcome the tidal dissipation that tends to break
  Cassini state 2 (eq.~[\ref{eqn:masslim}]).  The radial velocity dataset also constrains
  possible second planets; absence of a secondary signal of amplitude
  $K=5$~m~s$^{-1}$ puts an upper limit on mass at each orbital
  distance.  The arrows point to the portion of parameter space
  allowed by each particular constraint.  Apparently all of parameter
  space is ruled out: tides due to a high obliquity state, stabilized
  by the orbital precession induced by a second planet, cannot significantly
  inflate HD~209458b.  See \S\ref{sec:discussion} for derivation and
  details. }
\label{fig:masslim}
\end{figure}

  Alternatively, a second planet, rather than stellar obliquity, may
  be responsible for the orbital precession that has HD~209458b trapped in
  Cassini state 2 (see \S\ref{subsec:thirdbody}).  Here we compile the
  constraints and find that this hypothesis is also ruled out
  (Fig.~\ref{fig:masslim}).  Briefly, a putative second planet must be
  massive and distant enough to absorb the angular momentum of the
  dissipating planet, yet close enough so that its torque can compete
  with the dissipative torque, and these requirements would violate
  the radial velocity constraints.  Each of these constraints is
  explained next.
  
  How much angular momentum must be transfered to the second planet?
  If we happen to be observing the system near the end of Cassini
  state 2, such that it is nearly coplanar now, the final angular
  momentum in the orbit of HD~209458b is $L_f = L_{\rm now} = 1.26
  \times 10^{49}$~g~cm$^2$~s$^{-1}$.  We may combine the current
  inferred rate of orbital evolution with the expected scaling of
  $\dot{L}$ with $L$ to find its primordial value
  (\S\ref{subsec:thirdbody}).  Assuming Cassini state 2 was
  established at the inception of the system, the planet's primordial
  angular momentum was $L_i \approx 1.41 \times
  10^{49}$~g~cm$^2$~s$^{-1}$.  Thus the second planet must have enough
  non-aligned angular momentum such that $\Delta L_2 = 1.5 \times
  10^{48}$~g~cm$^2$~s$^{-1}$ is absorbed before coplanarity is
  established.  If we had assumed that the Cassini state was only
  midway through its evolution, this requirement would be even more
  strict.
  
 Using these values, we have that the primordial value for $i_{\rm
 rel,i}$ must have been greater than $i_{\rm crit} = 27^\circ$.  This
 value is considerably larger than relative inclinations of planets in
 our solar system, but it may be envisioned that planet-planet
 scattering, followed by planet-star tidal dissipation, was the cause
 of HD~209458b's current small semi-major axis, after which $i_{\rm
 rel}$ may plausibly be that large \citep{2002MW}.  On the other hand,
 for very large initial inclinations, eccentricity oscillations by the
 Kozai mechanism plus tidal friction will reduce $i_{\rm rel}$ to
 $\sim40^\circ$ before obliquity tides start to dominate dissipation,
 so it is reasonable to consider only initial values of $i_{\rm
 rel,i}$ less than $40^\circ$.

For each assumed primordial relative inclination, the second planet
must have at least the orbital angular momentum $L_2$ of
equation~(\ref{eqn:L2}) so that the two orbits are not yet coplanar.
This lower limit gives a lower limit to the mass of the putative
planet as a function of $a_2 (1-e_2^2)$, which we plot in
Figure~\ref{fig:masslim} for two values of $i_{\rm rel,i}$.

At large orbital separation, the torque of the second planet will not
be able to compete with the dissipative torques, and by analogy with
the stellar oblateness analysis, Cassini state 2 will not be an
equilibrium.  Here equation~(\ref{eqn:phisfo}) still applies, but
$\theta$ must be determined by setting the planet's spin precession
rate $g$ of equation~(\ref{eqn:gp}) equal to
\begin{equation}
g_2 = \frac34 \frac{G^{1/2} M_2}{M^{1/2}} \frac{a^{3/2} \cos i_{\rm rel,now}}{a_2^3 (1-e_2^2)^{3/2} },
\end{equation}
which is the orbital precession rate due to the second planet, whose
mass is $M_2$, semimajor axis $a_2$, eccentricity $e_2$, and current
relative inclination $i_{\rm rel,now}$.  Here $g_2$ is calculated by
secular theory, and any enhancement to this rate by the effects of
mean motion resonance will only be temporary, since $a$ is evolving
due to tidal dissipation. To find the critical parameters of a second
planet which barely compete with the dissipation torque, we may set
$\phi_s=1$ in equation~(\ref{eqn:phisfo}), and similar approximations
as those that led to
equations~(\ref{eqn:thetaapprox})-(\ref{eqn:edotmax}) give the
constraint
\begin{equation}
\frac{M_2}{a_2^3 (1-e_2^2)^{3/2}} > \frac{2 \times 4^{1/3}}{3} \dot{E}^{2/3} \frac{M_\star^{1/3}}{G} \frac{k_p^{1/3}}{C_p} 
\frac{R_p^{5/3}}{a} \frac{\cot^{2/3}{I} }{\cos i_{\rm rel,now} }. \label{eqn:masslim}
\end{equation}
If the exterior planet's orbit dominates the angular momentum, $i_{\rm
rel,now} \approx I$; the minimum value of $\cot^{2/3}{I} / \cos I$
occurs at $I = \tan^{-1} \sqrt{2} \approx 54.7^\circ$ and has the
value $3^{1/2}/2^{1/3} \approx 1.375$.  This angle is bigger than the
critical angle for Kozai eccentricity oscillations, so taking $1.375$
for the multiplicative factor depending on the inclinations is quite
conservative.  Then, evaluating the other terms with the values cited
for HD~209458b gives $M_2 a_2^{-3} (1-e_2^2)^{-3/2} > 3.74 M_{J}
\rm{AU}^{-3}$.  This constraint is labeled ``dissipative torque'' on
Figure~\ref{fig:masslim}.  It scales with $e_2$ differently than the
angular momentum constraint, but the limit shown is true for all
eccentricities.

The parameter space of possible second planets is also constrained by
the lack of a secondary signal in the radial velocity data, giving a
mass upper limit plotted in Figure~\ref{fig:masslim}.  No part of
parameter space remains, so the second planet mechanism for causing
the inflation of HD~209458b via Cassini state 2 is ruled out.

We have ruled out both the stellar rotational bulge and a second
planet as drivers of a high obliquity Cassini state for
HD~209458b, if it is presumed that obliquity tides are inflating its
radius.  We have focused on this planet as the observational material
far surpasses that of other planets.  However, we expect that the
hypothesis also fails for the other anomalously large planets
HAT--P-1b and WASP-1b.  Likewise, planetary systems discovered in the
future will need very peculiar parameters if obliquity tides are to be
an important heating source.

\section*{Acknowledgments} 
DF was supported by a NASA Origins of Solar Systems Program grant to S. Tremaine, and
EJ by NAG5-11664 to JG.  We thank Joshua Winn for detailed and helpful comments on a draft of this paper.

\bibliographystyle{apj}
\bibliography{FGJ07}
\clearpage


\begin{figure}
\center
\input{coords.pstex_t}
\caption{The $XYZ$ coordinate system precesses with angular frequency
  $g$ about the direction of the total angular momentum $\Jhat$.  The
  orbital plane, with normal $\hhat$ along the $Z$-axis, is inclined
  to the invariable plane by angle $I$ with the ascending node in the
  $X$ direction ($\Jhat$ lies in the $YZ$ plane).  The $xyz$ axes are
  the principal axes of the planet, with the unit spin vector $\Shat$
  along the axis of symmetry.  $\phi$, $\psi$, and the obliquity
  $\theta$ are Euler angles.  The phase angle
  $\phi_s\equiv\pi-\phi$.  Nondissipative Cassini states have
  $\Shat$ in the $YZ$ plane, with $\phi_s=\pi~(0)$ for state $1~(2)$;
  but dissipation causes $0<\phi_s\lesssim\pi/2$ in state 2.}
\label{fig:coords}
\end{figure}

\begin{figure}
\plotone{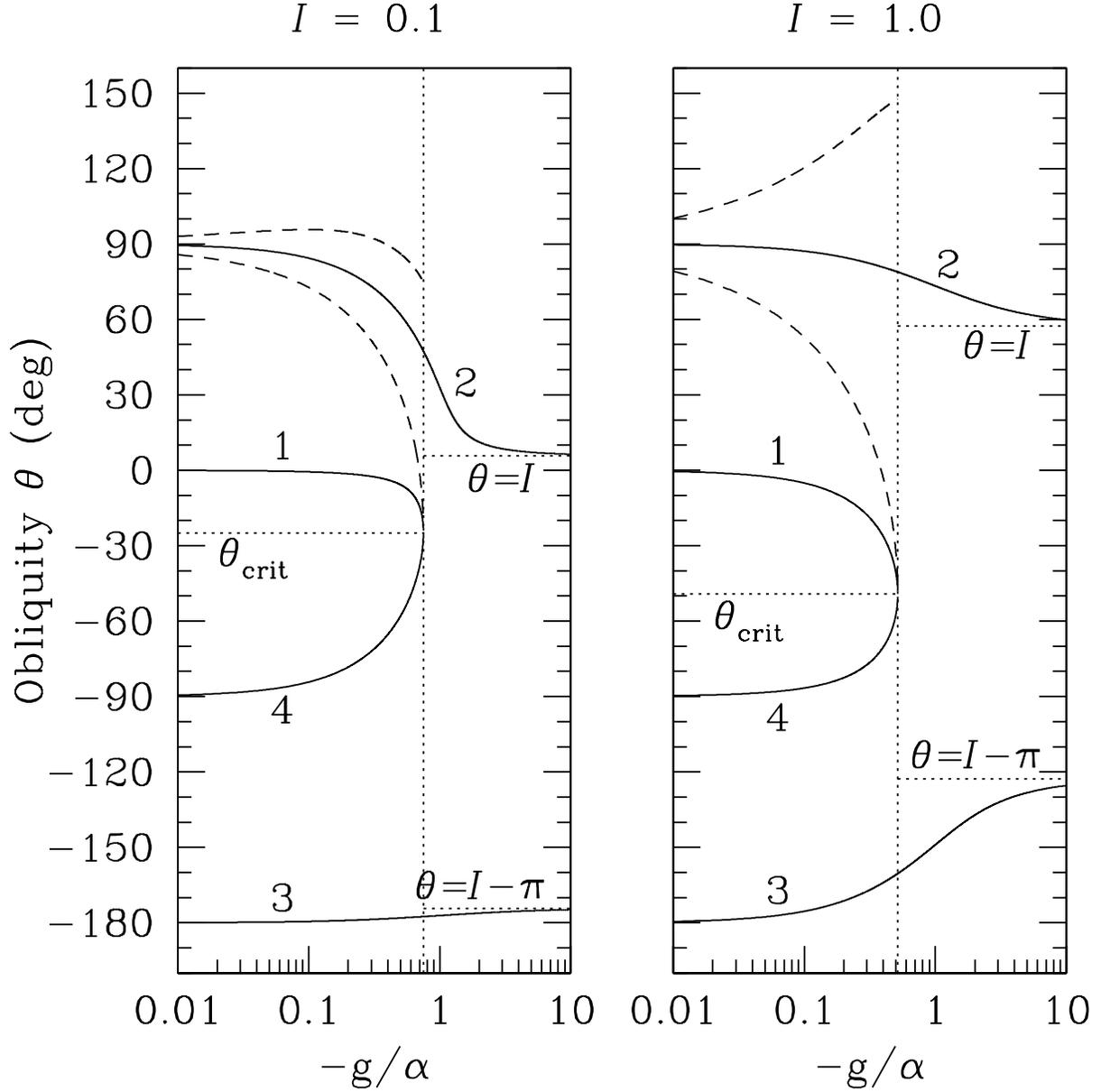}
\caption{Obliquities of the four Cassini states as functions of
$-g/\alpha$.  Solid numbered curves are the solutions of
equation~(\ref{Cassini}) for $I=0.1$ (left panel) and $I=1.0$ (right
panel).  Cassini states lie in the $YZ$ plane of Figure~\ref{fig:coords}; by convention $\theta$ is positive when $\phi_s=0$
and negative when $\phi_s=\pi$.  The separatrix intersects the $YZ$
plane at state 4, and at the angles indicated by the dashed curves.
States 1 and 4 merge and vanish at the critical value
$(-g/\alpha)_{\rm crit}=(\sin^{2/3}I+\cos^{2/3}I)^{-3/2}$ (dotted
vertical line).  Three values of $\theta$ are indicated by dotted
horizontal lines.  The largest absolute obliquity attained by state 1
is $\theta_{\rm crit}=\tan^{-1}(-\tan^{1/3}I)$.  States 2 and 3
approach $\theta=I$ and $\theta=I-\pi$, respectively, when
$|g/\alpha|\gg 1$.}
\label{fig:obliq}
\end{figure}

\begin{figure}
\plotone{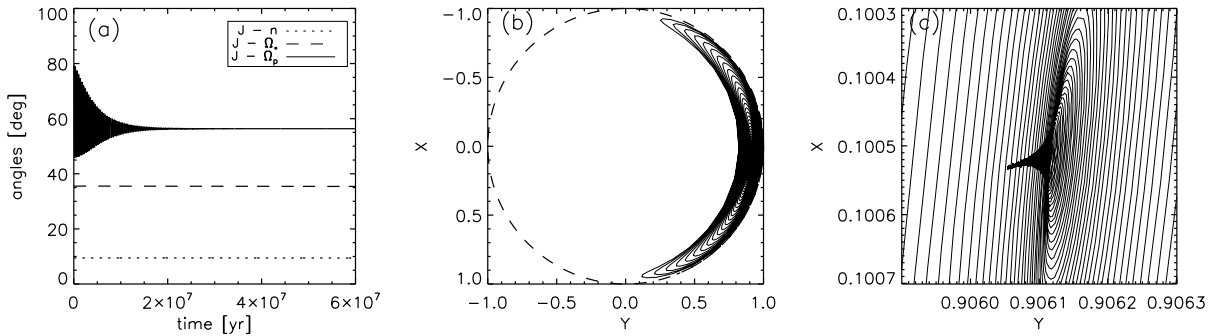}
\caption{ Settling into Cassini state 2.  (a) The angles between
${\bf J}$ and the other vectors as a function of time. (b) The motion
of ${\bf \hat \Omega_p}$ on the unit sphere, in the coordinate system
of Figure~\ref{fig:coords}.  (c) After the equilibrium state is
reached, the system slowly evolves (see Fig.~\ref{fig:cassinilong}).
Notice that the state does not have equilibrium value $\phi =
\pi$, as expected from the analysis with no dissipation.  In this case
there is a phase shift away from the Y axis of $\phi_s = 0.1105$.  We have used an 
exceptionally low value of the apsidal motion constant $k_p = 5 \times 10^{-5}$ for this integration (see text).}
\label{fig:state2settle}
\end{figure}

\begin{figure}
\plotone{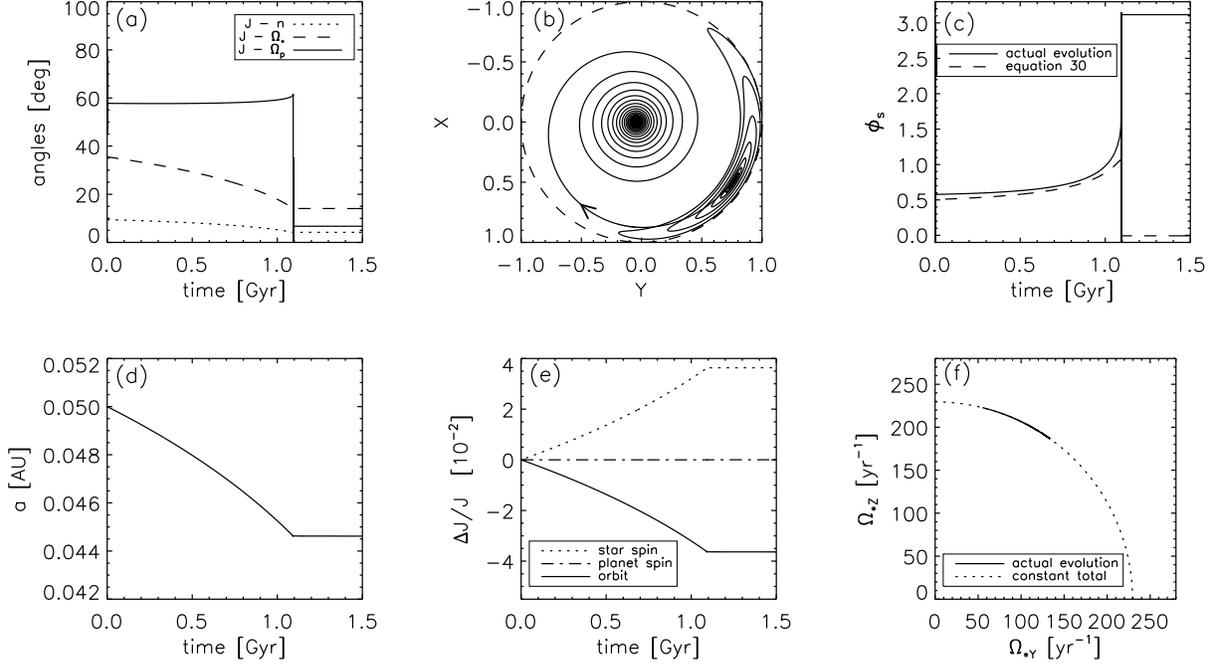}
\caption{ The long term evolution of Cassini state 2.  (a) The angles
between ${\bf J}$ and the other vectors as a function of time. (b) The
motion of ${\bf \hat \Omega_p}$ on the unit sphere, in the coordinate
system of Figure~\ref{fig:coords}.  This longer integration shows that after damping into Cassini state 2, as in Figure~\ref{fig:state2settle}(b), the planetary spin vector evolves to negative Y, after which equilibrium is lost it librates around, and damps into, Cassini state 1.  (c) The evolution of the phase shift angle
$\phi_s \equiv \pi - \phi$.  Most of the time in the parametric plot of panel (b) is spent in  the evolving state 2; libration epochs are too short to be resolved on this plot.  The dashed line is the first-order estimate of
equation~(\ref{eqn:phisfo}).  (d) The obliquity tides cause the semi-major
axis to shrink.  After damping to Cassini state 1, no appreciable energy is lost.  (e) The changes in the projections of the three
angular momenta along the $\Jhat$ axis, as a function of time.  The
angular momentum lost by the planet as it migrates in is absorbed by
the stellar spin.  (f) The stellar spin absorbs the angular momentum
of the planetary orbit by reorienting toward alignment.  The dashed
line is the set of points with equal magnitude of $\Omega_\star$,
showing that the only change to the stellar spin is its direction.  We have used an exceptionally low value of the apsidal motion constant $k_p = 5 \times 10^{-5}$ for this integration (see text). }
\label{fig:cassinilong}
\end{figure}

\end{document}